\documentclass[]{aa}
\usepackage{natbib}         
\usepackage{graphicx}
\usepackage{longtable}
\usepackage{xcolor}
\usepackage[switch]{lineno} 
\bibpunct{(}{)}{;}{a}{}{,} 
\def\nombre{372} 

\newcommand{\ind}[1]{_{\mathrm{#1}}}
\newcommand{\diff}{\mathrm{d}}

\newcommand\Dnu{\Delta\nu}
\newcommand\Dnup{\Delta\nu\ind{p}}
\newcommand\Tg{\Delta\Pi_1}
\newcommand\Teff{T\ind{eff}}
\newcommand\dens{\mathcal{N}}
\newcommand\dfres{\delta\! f\ind{res}}

\newcommand\dnurot{\delta\nu\ind{rot}}

\newcommand\dnurotcore{\delta\nu\ind{rot,core}}

\newcommand\nup{\nu\ind{p}}\newcommand\nug{\nu\ind{g}}

\newcommand\numax{\nu\ind{max}}
\newcommand\nmax{n\ind{max}}

\newcommand\DP{\Delta P}

\newcommand\thetap{\theta\ind{p}}
\newcommand\thetagg{\theta\ind{g}}
\newcommand\np{{n\ind{p}}}\renewcommand\ng{{n\ind{g}}}
\newcommand\nm{{n}}
\newcommand\epsp{\varepsilon\ind{p}}
\newcommand\epsg{\varepsilon\ind{g}}
\newcommand{\BV}{Brunt-V\"ais\"al\"a}

\newcommand{\zmoyn}{\langle\zeta\rangle_\nm}
\newcommand{\zmoym}{\langle\zeta\rangle_m}
\newcommand{\Rp}{R_p}
\newcommand{\Tobs}{T\ind{obs}}
\newcommand{\Dlarge}{\Delta_\nu}

\newcommand\Kepler{\emph{Kepler}}

\begin{document}

\title{Period spacings in red giants\footnote{Table C.1 is only available in electronic form
at the CDS via anonymous ftp to cdsarc.u-strasbg.fr (130.79.128.5)
or via \texttt{http://cdsweb.u-strasbg.fr/cgi-bin/qcat?J/A+A/}}}
\subtitle{IV. Toward a complete description of the mixed-mode pattern}
\titlerunning{Mixed modes}
\author{%
 B. Mosser\inst{1},
 C. Gehan\inst{1},
 K. Belkacem\inst{1},
 R. Samadi\inst{1},
 E. Michel\inst{1},
 M-J. Goupil\inst{1}
}

\institute{
 LESIA, Observatoire de Paris, PSL Research University, CNRS, Universit\'e Pierre et Marie Curie,
 Universit\'e Paris Diderot,  92195 Meudon, France; \texttt{benoit.mosser@obspm.fr}
 }


\abstract{Oscillation modes with a mixed character, as observed in evolved low-mass stars, are highly sensitive to the physical properties of the innermost regions. Measuring their properties is therefore extremely important to probe the core, but requires some care, due to the complexity of the mixed-mode pattern. }
{The aim of this work is to provide a consistent description of the mixed-mode pattern of low-mass stars, based on the asymptotic expansion. We also study the variation  of the gravity offset $\epsg$ with stellar evolution.}
{We revisit previous works about mixed modes in red giants and empirically test how period spacings, rotational splittings, mixed-mode widths, and heights can be estimated in a consistent view, based on the properties of the mode inertia ratios.}
{From the asymptotic fit of the mixed-mode pattern of a large set of red giants at various evolutionary stages, we derive unbiased and precise asymptotic parameters. As the asymptotic expansion of gravity modes is verified with a precision close to the frequency resolution for stars on the red giant branch (10$^{-4}$ in relative values), we can derive accurate values of the asymptotic parameters. We decipher the complex pattern in a rapidly rotating star, and explain how asymmetrical splittings can be inferred. We also revisit the stellar inclinations in two open clusters, NGC 6819 and NGC 6791: our results show that the stellar inclinations in these clusters do not have privileged orientation in the sky. The variation of the asymptotic gravity offset with stellar evolution is investigated in detail. We also derive generic properties that explain under which conditions mixed modes can be observed.}
{}


\maketitle

\section{Introduction}

Probing the cores of stars is difficult since, generally,
stellar information arises from their photosphere. Fortunately,
asteroseismology of evolved stars reveals stellar interiors in a
unique and powerful way: gravity waves that propagate throughout
the core couple with pressure waves and construct mixed modes that
can be observed \citep{2011Sci...332..205B,2011Natur.471..608B,2014ApJ...781L..29B}. The measurement of the global seismic properties of these mixed modes then carries unique information on the core structure
\citep[e.g.,][]{2013ApJ...766..118M,2016MNRAS.457L..59L,2015MNRAS.453.2290B,2017MNRAS.469.4718B}.
Observations with the space missions CoRoT and \emph{Kepler} have
provided the measurement of the asymptotic period spacings
\citep{2012A&A...540A.143M,2016A&A...588A..87V}, of the differential-rotation profile
in red giants \citep{2012Natur.481...55B,2014A&A...564A..27D,2015A&A...580A..96D}, and of the core
rotation for about 300 stars analyzed by \cite{2012A&A...548A..10M}.

Most of the previous studies are based on the measurement and analysis of global seismic parameters, such as the asymptotic large separation $\Dnu$ and the asymptotic period spacings $\Tg$
\citep[e.g.,][]{2017AN....338..644M}.
It is now time to access the properties of individual frequencies
in red giants. Up to now, most of the studies
\citep[e.g.,][]{2012A&A...538A..73B,2016ApJ...817...65D} were limited to stars on the red giant branch (RGB). Two main reasons explain this restriction: first, the oscillation spectra benefit from a better relative frequency resolution for this evolutionary stage; second, the oscillation spectra remain simple, with rotational splittings smaller than period spacings. When stars evolve, these features
become intricate, so that confusion is possible. For the most evolved stars, mixed modes are no longer observable \citep[e.g.,][]{2012A&A...538A..73B,2013A&A...559A.137M,2014ApJ...788L..10S}.

The understanding of any complicated mixed-mode oscillation pattern must be based on an unambiguous identification of the modes. Up to now, the most efficient method has relied on the use of the
asymptotic expansion, completed by a clear description of the influence of rotation \citep{2015A&A...584A..50M}. New insights on rotation were provided by the analysis depicted in
\cite{2016arXiv161205414G}, who have developed a methodology to measure rotational splittings in an automated way; \cite{2017EPJWC.16004005G} and \cite{2018arXiv180204558G} showed how rapid rotation can be
addressed efficiently. This efficiency derives from the use of stretched oscillation spectra.

In this work, we first examine in Section \ref{parametres} how the different frequency spacings in the asymptotic mixed-mode expansion can be expressed as a function of the mode inertia. New expressions are proposed for the mixed-mode spacings and rotational splittings. Case studies are examined in Section \ref{casestudy} to test and validate these expressions. In Section \ref{application-asymptotic}, we take advantage of the precision of the fits to derive accurate asymptotic period spacings and gravity offsets; for the first time, we can exhibit the global evolution of these gravity offsets as a function of stellar evolution. New insights on the rotational splittings are proposed in Section \ref{application-rotation}; in particular, we show how the asymptotic expansion can be used to provide priors based upon physical assumptions for any fitting code used later in the analysis. Finally, we assess the  conditions for observing mixed modes, based on global asymptotic parameters only (Section \ref{application-observability}). Section \ref{conclusion} is devoted to our conclusions.

\section{Mixed-mode parameters\label{parametres}}

Following the work of \cite{1979PASJ...31...87S} and \cite{1989nos..book.....U}, we derived  asymptotic expansions of mixed modes for different seismic parameters: eigenfrequencies \citep{2012A&A...540A.143M}, period spacings \citep{2012ASPC..462..503C}, rotational splittings \citep{2013A&A...549A..75G,2015A&A...580A..96D}, and mode widths and mode heights
\citep{2014A&A...572A..11G,2015A&A...579A..30B,2015A&A...579A..31B,2017A&A...598A..62M}. Here, we intend to revisit all these parameters that depict the mixed-mode spectrum in order to provide a more precise and unified view.

\subsection{Asymptotic expansion}

The asymptotic expansion of mixed modes is an implicit relation between the phases $\thetap$ and $\thetagg$ of the pressure- and gravity-wave contributions to the mixed modes, respectively. It reads
\begin{equation}\label{eqt-asymp}
    \tan\thetap = q \tan\thetagg ,
\end{equation}
where $q$ is the coupling factor \citep{2017A&A...600A...1M}. The phases are related to the large separation $\Dnu$ and the period spacing $\Tg$. The most convenient expressions of the phase refer respectively to the pure\footnote{Pure p (or g) modes are hypothetical modes that could be formed in the pressure (or gravity) cavity without any coupling with the other cavity.} p and g mode spectra
\begin{eqnarray}
  \thetagg &=& \pi {1 \over \Tg}  \left({\displaystyle{1\over\nu}
  -\displaystyle{1\over\nug}}\right), \label{eqt-g} \\
  \thetap &=& \pi {\nu-\nup\over \Dnup} \label{eqt-p},
\end{eqnarray}
where $\nup$ and  $\nug$ are the asymptotic frequencies of pure pressure and gravity modes, respectively, and $\Dnup$ is the frequency difference between the consecutive pure pressure radial modes with radial orders $\np$ and $\np+1$. In this work, we consider that the radial modes and pure dipole pressure modes obey the universal red giant oscillation pattern \citep{2011A&A...525L...9M} and that the dipole gravity modes follow the asymptotic comb-like pattern
\begin{equation}\label{eqt-asymp-g}
  {1\over \nug} = (-\ng + \epsg) \ \Tg
  ,
\end{equation}
where $\Tg$ is the period spacing and $\epsg$ is the gravity offset.

\cite{2015A&A...584A..50M} derived that the variation of the
oscillation period $P$ with the mixed radial order $\nm$ writes
\begin{equation}\label{eqt-derivee}
    {\diff P \over \diff\nm} = \zeta \ \Tg
    .
\end{equation}
A convenient way to write the parameter $\zeta$ is
\citep{2017A&ARv..25....1H}
\begin{equation}\label{eqt-zeta}
    \zeta(\nu) = \left[1+ {q\over \dens }
     {1 \over q^2 \cos^2\thetap + \sin^2 \thetap}
    \right]^{-1}
    ,
\end{equation}
where $\dens =\Dnu/(\nu^2 \Tg)$ is the density of gravity modes
compared to pressure modes, in other words the number of mixed modes in a $\Dnu$-wide interval. Compared to the original form presented in \cite{2015A&A...584A..50M}, the rapidly varying phase $\thetagg$ has been replaced by a function of $\thetap$ that varies in a smooth way.

As demonstrated by \cite{2013A&A...549A..75G} and used by subsequent work \citep{2014ApJ...781L..29B,2015A&A...580A..96D}, the function $\zeta$ is connected to the 
inertia of mixed modes. Introducing the contributions of the envelope and of the core, \begin{equation}\label{eqt-zeta-inerties-def}
   \zeta = {I\ind{core} \over I\ind{env}+ I\ind{core}}
   ,
\end{equation}
and assuming that the envelope contribution of a mixed mode is similar to the inertia of the closest radial mode ($I\ind{env} \equiv I_{\np,0}$),  we find that the inertia of the dipole mode with mixed
radial order $\nm$ varies  as
\begin{equation}\label{eqt-zeta-inerties}
    I_{\nm,1} = { I_{\np,0} \over 1 - \zeta}
    .
\end{equation}
For the sake of simplicity, we use hereafter the abridged notation $I_\nm$ for the inertia of the dipole mixed modes and $I_0$ for the closest radial modes, and follow the same convention for the mode heights and widths.

\subsection{Seismic parameters}

With $\zeta$, we now intend to express the different seismic
parameters.

\subsubsection{Period spacing}

Following  \cite{2012ASPC..462..503C} and \cite{2015A&A...584A..50M}, period spacings can be expressed as
\begin{equation}\label{eqt-zeta-P0}
    \DP = P_\nm - P_{\nm+1} =
     \zeta\ \Tg
     .
\end{equation}
This expression is however ambiguous, since $\zeta$ may vary significantly between the periods $P_{\nm+1}$ and $P_\nm$ ($> P_{\nm+1}$). Therefore, we prefer to consider the expression resulting from the integration of
Eq.~(\ref{eqt-derivee})
\begin{equation}\label{eqt-zeta-P}
    \DP = P_\nm - P_{\nm+1} =
    \Tg\ \int_{\nm}^{\nm+1} \zeta(\nu) \ \diff n
    = \Tg\ \zmoyn
    ,
\end{equation}
where we consider that the mixed-mode radial order $n$ is a continuous variable
defined by $\diff n = \diff \tau / \Tg$, where $\tau$ is the stretched period introduced by \cite{2015A&A...584A..50M}; i.e.,
\begin{equation}\label{eqt-stretched}
 \diff \tau = {\diff \nu \over \zeta\ \nu^2}
  .
\end{equation}
In fact, $n$ takes consecutive integer values for each mixed mode. In this work, we use an estimate of $n = \np + \ng$ derived from the pressure and gravity radial orders; $\np$ is derived from the universal red giant oscillation pattern \citep{2011A&A...525L...9M}, whereas $\ng$ is given by
\begin{eqnarray}
  \ng = - \left\lfloor{{1\over \nu \,\Tg} - {1\over 4}}\right\rfloor \hbox{ \ on the RGB,}\label{eqt-estim-ng}\\
  \ng = - \left\lfloor{{1\over \nu \,\Tg} + {1\over 4}}\right\rfloor   \hbox{ \ in the red clump},
\end{eqnarray}
where the correcting terms $\pm 1/4$ that depend on the evolutionary stage are justified in Section~\ref{gravity-offsets}. They differ by 1/2, as depicted by the asymptotic relation \citep[e.g.,][]{1980ApJS...43..469T,2013ApJ...767..158B}. In red giants, the high density $\dens$ of mixed modes implies that $|\ng| \gg \np$, so that the mixed-mode orders are negative.

From the definition of the stretched period, Eq.~(\ref{eqt-zeta-P}) reduces to
\begin{equation}\label{eqt-zeta-P-int}
    \DP
    =
    \int_{\nu_\nm}^{\nu_{\nm+1}} { \diff \nu \over \nu^2}
    .
\end{equation}
This evident relation justifies the relevance of
Eq.~(\ref{eqt-zeta-P}) instead of Eq.~(\ref{eqt-zeta-P0}): using $\zmoyn$ is necessarily more accurate than using $\zeta$ for computing period spacings.

\subsubsection{Rotational splitting}

As introduced by \cite{2013A&A...549A..75G}, the function $\zeta$
is used to express the mixed-mode rotational splitting as a
function of the mean rotational splittings related to pure gravity
or pure pressure modes:
\begin{equation}\label{eqt-zeta-rotation}
     \dnurot =
     \zeta \ {\dnurot}\ind{,g}
     +
     (1-\zeta) \ {\dnurot}\ind{,p}
     .
\end{equation}
As shown by subsequent works \citep[e.g.,][]{2014A&A...564A..27D,2016ApJ...817...65D,2017A&A...602A..62T},
it is difficult to derive from the observed rotational splittings
more than these two mean quantities.

Again, we have to solve the ambiguity of the meaning of $\zeta$ in
Eq.~(\ref{eqt-zeta-rotation}), since we can either consider the
value\footnote{Since we consider dipole modes only, we use a simplified notation $\nu_{n,m}$ instead of $\nu_{n,\ell,m}$.} $\zeta(\nu_{\nm,0})$, in the framework of the perturbation
of the non-rotating frequency $\nu_{\nm,0}$, or $\zeta(\nu_{\nm,m})$, considering that the inertia to be
considered corresponds to the actual frequency $\nu_{\nm,m}$.
By analogy with the equation dealing with the period spacing, we propose to rewrite the rotational splitting $\dnurot =  \nu_{\nm,m} - \nu_{\nm,0}$, in the limit case where the mean envelope
rotation is negligible compared to the mean core rotation, as
\begin{equation}\label{eqt-zeta-rotation2}
  \dnurot
  = \dnurotcore \int_{\nu_{\nm,0}}^{\nu_{\nm,m}} \zeta\ \diff m
  = \dnurotcore \   \zmoym
  ,
\end{equation}
where $\dnurotcore \equiv {\dnurot}\ind{,g}$. As for the radial order $\nm$ in Eq.~(\ref{eqt-zeta-P}), we consider the azimuthal order $m$ as a continuous variable varying from 0 to $\pm 1$. So, we have introduced two mean values of $\zeta$,
\begin{eqnarray}
  \zmoyn &=& \int_{\nm}^{\nm+1} \zeta \ \diff n
  = \int_{\nu_{\nm,m}}^{\nu_{\nm+1,m}} { \diff \nu \over \Tg \,\nu^2},
  \label{eqt-zmoy-n} \\
  \zmoym  &=& \int_0^{\pm 1} \zeta\ \diff m
  = \int_{\nu_{\nm,0}}^{\nu_{\nm,\pm 1}}   { \diff \nu \over \Tg \, \nu^2}
  ,
  \label{eqt-zmoy-m}
\end{eqnarray}
to account for the period spacings and rotational splittings. The relevance of $\zmoyn$ is already proven by Eqs.~(\ref{eqt-zeta-P}) and (\ref{eqt-stretched}), whereas the relevance of $\zmoym$ has yet to be demonstrated. If we succeed, we will also have understood the relevance of the use of stretched periods for analyzing the mixed modes (Eq.~\ref{eqt-stretched}).

\subsection{Mixed-mode width, height, and amplitude}

The work performed by the gas during one oscillation cycle  is the same for all modes, associated with surface damping, when the radiative damping in the \BV\ cavity is considered as negligible. Hence,  \cite{2014ApJ...781L..29B} have estimated that the mode width of the mixed modes writes
\begin{equation}\label{eqt-largeur}
  \Gamma_\nm = \Gamma_0 \ {I_0 \over I_\nm}
   = \Gamma_0 \ (1 - \zeta)
   .
\end{equation}
From this relation, we verify that mixed modes have smaller mode
widths than radial modes. However, we recall that a family of stars behave differently, when mixed modes are depressed because of an extra damping in the radiative inner region \citep{2012A&A...537A..30M,2014A&A...563A..84G,2017A&A...598A..62M}.

From \cite{2015A&A...579A..31B} we also derive that the amplitude
of a resolved dipole mixed mode is
\begin{equation}\label{eqt-zeta-a2}
     A^2_\nm = A_0^2\ (1-\zeta)
     ,
\end{equation}
when the geometrical factor that conducts to a visibility of
about 1.54 for red giant dipole modes
\citep{2012A&A...537A..30M,2017A&A...598A..62M} is omitted. Such amplitudes correspond to similar heights for radial and dipole modes since $A^2 = \pi\Gamma H / 2$. When, for non-resolved mixed modes, the width $\Gamma_\nm$ is less than the frequency resolution $\dfres$, a dilution factor must be considered \citep{2009A&A...506...57D}. It expresses
\begin{equation}\label{eqt-zeta-H}
     H_\nm =  {\pi \over 2} \ {\Gamma_\nm \over \dfres} \ H_0
     ,
\end{equation}
when radial modes are resolved, which is the common case.

\begin{figure}
\includegraphics[width=8.8cm]{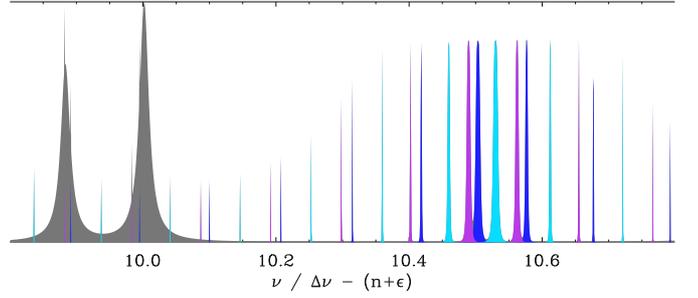}
\caption{Description of the radial order closest to $\numax$ of the oscillation power spectrum of a typical RGB star. Frequencies, widths, and heights are estimated according to the function $\zeta$. Quadupole and radial modes are plotted in gray, dipole mixed modes in dark blue ($m=-1$), light blue ($m=0$), or purple ($m=1$), respectively.}\label{fig-asymetrie}
\end{figure}

\subsection{Synthetic mixed-mode pattern}

The previous ingredients can be used to depict an oscillation
pattern. Figure~\ref{fig-asymetrie} shows the synthetic spectrum
of a typical star on the low RGB, based on Eq.~(\ref{eqt-zeta-rotation2}) for the rotational splittings, on Eq.~(\ref{eqt-largeur}) for the mode widths, and on Eq.~(\ref{eqt-zeta-H}) for the mode heights of unresolved modes. This spectrum resembles the description derived by \cite{2014A&A...572A..11G} from non-adiabatic computations, with a time-dependent treatment of convection which provides the lifetimes of radial and non-radial mixed modes.

\begin{figure*}
\includegraphics[width=15cm]{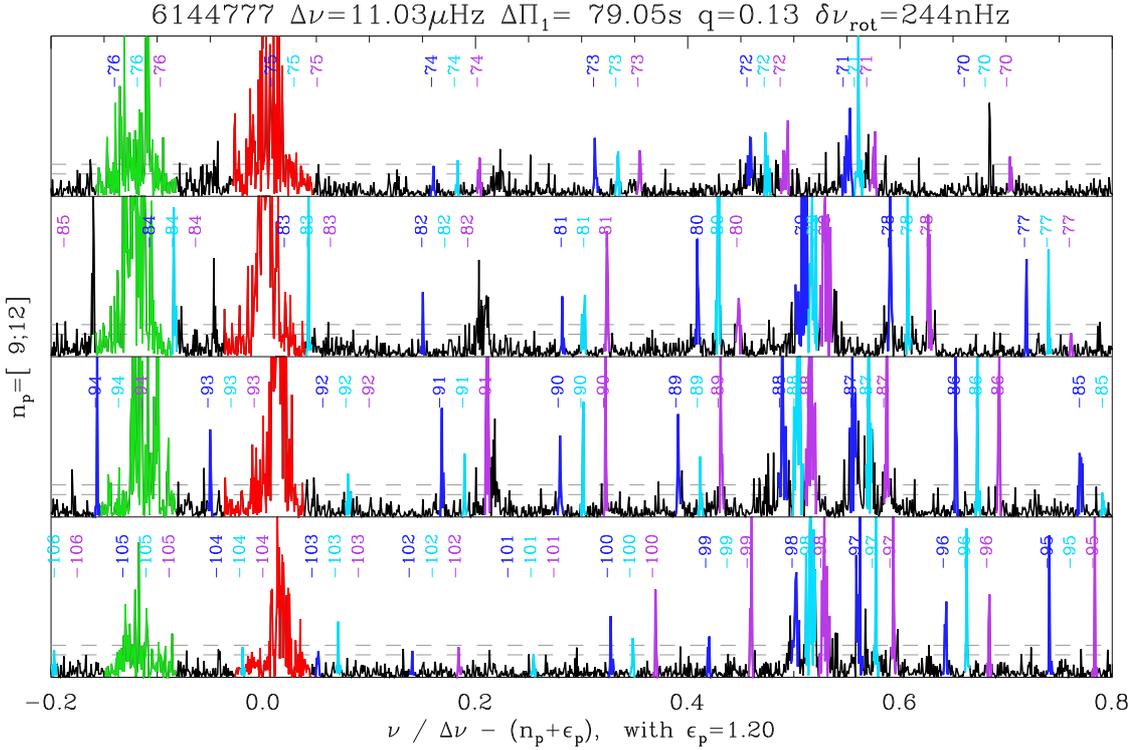}
\caption{Fit of the oscillation pattern of the RGB star KIC 6144777, showing the pressure radial orders $\np$ from 9 to 12. The power spectrum density has been divided by the fit of the background. Radial and quadrupole modes are highlighted in red and green. The expected locations of dipole mixed modes are labelled with their mixed radial orders. When detected, mixed modes are highlighted in dark blue ($m=-1$), light blue ($m=0$), or purple ($m=1$). $\ell=3$ modes, which are also mixed, are located near the abscissa 0.22; extra peaks in the range $[-0.2,\,-0.05]$ are mixed quadrupole modes. The gray dashed lines indicate the two thresholds used in this work, corresponding to height-to-background ratios of 7 and 10.
}\label{fig-fit-6144777}
\end{figure*}

\section{Case studies\label{casestudy}}

In this Section, we use RGB stars showing clear oscillation spectra as case studies, in order to test the description of the mixed-mode spacings, widths, heights, and rotational splittings, which were previously introduced. The first steps consist in identifying their oscillation spectra and in fitting as many dipole mixed modes as possible. One of the two stars considered here, KIC 6144777 was already investigated in many previous articles \citep[e.g.,][]{2015A&A...579A..83C,2018MNRAS.476.1470G}. The other one, KIC 3955033, was less studied since it shows a complicated mixed-mode spectrum; it belongs to the list of red giants with period spacings automatically computed by \cite{2016A&A...588A..87V}. We used data downloaded from the KASOC site\footnote{http://kasoc.phys.au.dk}, processed using the \Kepler\ pipeline developed by \cite{2010ApJ...713L..87J}, and corrected from outliers, occasional jumps, and drifts \citep[see][for details]{2011MNRAS.414L...6G}.

\subsection{Identification of the mixed modes\label{identi_mm}}

The location of the mixed modes primarily relies on the firm identification of the pure pressure-mode spectrum. The determination of the large separation $\Dnu$, first derived from the envelope autocorrelation function \citep{2009A&A...508..877M}, is based on the universal red giant oscillation pattern. This method provides the efficient identification of the radial modes and helps to locate the frequency ranges where mixed modes cannot be mistaken for radial or quadrupole modes. For $\ell=1$ modes, the second-order asymptotic expansion writes
\begin{equation}\label{eqt-asymp01}
   \nup = \left(\np  + \varepsilon\ind{p} + {1\over 2} + d_{01}
   + {\alpha \over 2}\; [ \np - \nmax ]^2 \right) \; \Dnu
   ,
\end{equation}
where $ \varepsilon\ind{p}$ is the acoustic offset, $\nmax = \numax/\Dnu -  \varepsilon\ind{p}$, and $\alpha=0.076/\nmax$. The parameter $d_{01}$ is function of the large separation, under the form
$A+ B \log\Dnu$ (where $\Dnu$ is expressed in $\mu$Hz), with $A = 0.0553$ and $B=-0.036$, as determined from the large-scale analysis along the RGB conducted by \cite{2014A&A...572L...5M}. The accurate determination of $d_{01}$ is crucial for the determination of the pure dipole pressure modes, hence for the determination of the minima of the function $\zeta$. In that respect, the small modulation of the radial-mode pattern induced by the sound-speed glitches \citep{2010A&A...520L...6M,2015A&A...579A..84V} must be considered also. Therefore, we fit the actual position of the radial modes first, then use them to refine the pure pressure dipole-mode frequencies according to
\begin{equation}\label{eqt-asymp1}
   \nup =\left( {\nu_{\np,0} + \nu_{\np+1,0}} \right) / 2 + d_{01}  \; \left( {\nu_{\np+1,0} - \nu_{\np,0}} \right)
   .
\end{equation}
The background parameters, derived as in \cite{2012A&A...537A..30M}, are used to correct the granulation
contribution in the frequency range around $\numax$. Hence, mixed modes can be automatically identified in frequency ranges that have no radial and quadrupole modes when their heights are significantly above the background. The automatic selection of the modes relies on a statistical test: the height-to-background ratio of the modes must be higher than a threshold level $\Rp$ in order to reject the null hypothesis to a low probability $p$. According to \cite{2006ESASP1306..377A}, the relation between $\Rp$ and  $p$ depends for long-lived modes on the observation duration $\Tobs$ and on the width $\Dlarge$ of the frequency range where a mode is expected. This relation expresses
\begin{equation}\label{eqt-seuil}
  \Rp \simeq \ln{\Tobs \Dlarge \over p},
\end{equation}
when expressed in noise unit. This situation applies here, since the precise identification of the mixed-mode pattern is based on gravity-dominated mixed modes. With 4-year observations and the search of a couple of modes in a frequency range $\Dlarge = \Dnu / \dens$, the threshold is typically 10 for a secure probability rejection at the $10^{-2}$ level. In practice, mixed modes with a height-to-background ratio higher than 10 are used to initiate the fit. A lower threshold is enough for the final agreement, when the synthetic mixed-mode pattern based on secure modes can be used to search for long-lived mixed modes in narrow frequency ranges.
We benefit from the fact that the asymptotic fit is precise and enables to search for thin modes in a frequency range $\Dlarge$ narrower than 0.1\,$\mu$Hz. Therefore, a threshold of 7 is enough for rejecting the null hypothesis at the 1\,{\%}-level for these modes whose detection benefits from the information gained by larger peaks. The thin mode widths (Eq.~\ref{eqt-largeur}) are of great use to map the observed spectrum: a thin gravity-dominated mixed mode must be found in the close vicinity, less than 4 times the mode width, of its expected position. For unresolved peaks, this condition is relaxed to 4 times the frequency resolution. The global seismic parameters of the gravity component are then derived from the methods described in \cite{2016A&A...588A..87V} and \cite{2017A&A...600A...1M}, with a least-square fit between the observed and asymptotic patterns.

At this stage, global seismic parameters are measured and mixed modes are identified, so that it is possible to measure their individual properties.

\begin{figure}
\includegraphics[width=9.0cm]{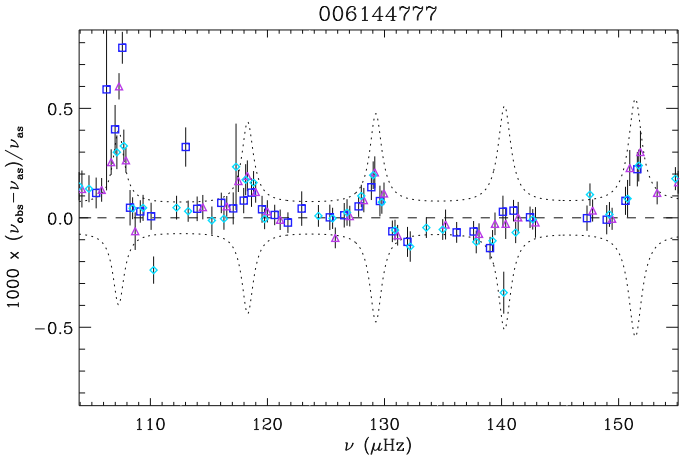}
\caption{Relative residuals, multiplied by 1000, between the observed and asymptotic mixed-mode frequencies in KIC 6144777. The color of the symbols indicates the azimuthal order: dark blue squares for $m=-1$, light blue diamonds for $m=0$, or purple triangles for $m=1$; 1-$\sigma$ uncertainties are also shown. The dashed line corresponds to a perfect fit. The dotted lines show the frequency resolution plus an extra-modulation $\Dnu (1-\zeta) /100$, which is empirically used to define the quality of the fit.}
\label{fig-res-6144777}
\end{figure}

\subsection{Individual fitting procedure}

When fitting individually mixed modes, we aim at testing the validity of
the asymptotic expression, but not at reaching the ultimate precision, which is the role of a dedicated fit of individual modes \citep[e.g.,][]{2009A&A...506....7G}. Therefore, in order to
simplify the fit, we supposed (and checked a posteriori) that all
multiplets can be fitted independently. This is not the
case in all red giant spectra, but it is verified for most stars
on the early RGB or in the red clump.

From the asymptotic fit, we identify in the background-corrected
spectrum the power excess associated to each mode. Then, we determine the central frequency of the peak as the barycenter of the power excess. The height $H$ and full width at half maximum $\Gamma$ are simultaneously derived from the Lorentzian fit of the mode. We use Eqs.~(\ref{eqt-largeur}) and (\ref{eqt-zeta-H}) as priors. Modes are fitted individually when the mode density is low, or simultaneously when the Lorentzians used as priors overlap.

The fitted spectrum  and the seismic parameters of KIC 6144777, used as a first study case, are given in Fig.~\ref{fig-fit-6144777} and in Table~\ref{table-freq6144777}. We note the large agreement between the observed and asymptotic peaks. As in other stars showing a seismic signal with a high signal-to-noise ratio (S/R), outliers with a height-to-background value $R$ higher than 7 are present. Their detection does not invalidate the method presented above: they correspond either to $\ell=2$ or 3 mixed modes, possibly also to $\ell=4$ modes, or to aliases (since the duty cycle is about 93\,\%), or even to noise since the detection of 1 noisy peak with $R\ge8$ is expected in a 30-$\mu$Hz frequency range after 4 years of observation, assuming that the noise statistic is a $\chi^2$ with two degrees of freedom.

The quality of the fit is shown by the small residuals between the observed frequencies and the asymptotic fits (Fig.~\ref{fig-res-6144777}); we note that these residuals are comparable to the uncertainties, derived from \cite{1992ApJ...387..712L} or slightly larger when the quality of the fit may be affected by the high mode density. These residuals  are of about the frequency resolution. When pressure-dominated mixed modes are excluded, the standard deviation of the asymptotic fit is 11\,nHz. This value represents 1.3 times the frequency resolution $\dfres$, or $\Dnu/1000$, or a relative precision at $\numax$ of about $10^{-4}$. The quality of the fits is based on a small number of parameters:  the radial mode frequencies, the mean location $d_{01}$ of the expected pure pressure dipole modes, and four asymptotic parameters: the period spacing $\Tg$, the coupling factor $q$, gravitational offset $\epsg$, and the mean core rotational splitting $\dnurot$. Residuals reach maximum values near the pressure-dominated mixed modes: there, deviations of about $\Dnu/200$ are observed, to be compared to the amplitudes of pressure glitches of about $\Dnu/40$ \citep{2015A&A...579A..84V}. We suspect that these residuals are mostly due to the variation of the parameter $d_{01}$ with frequency.

\begin{figure}
\includegraphics[width=8.8cm]{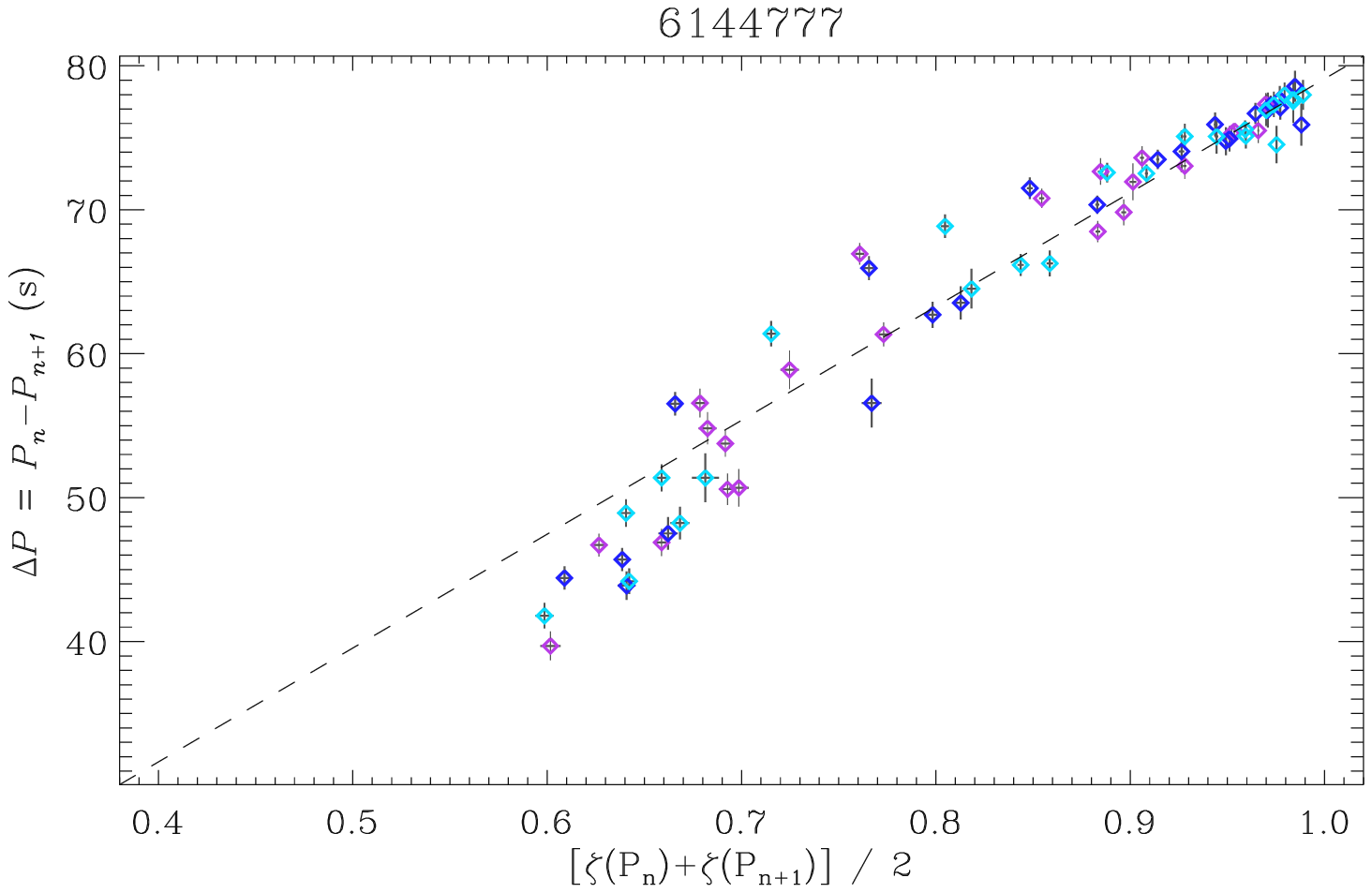}
\includegraphics[width=8.8cm]{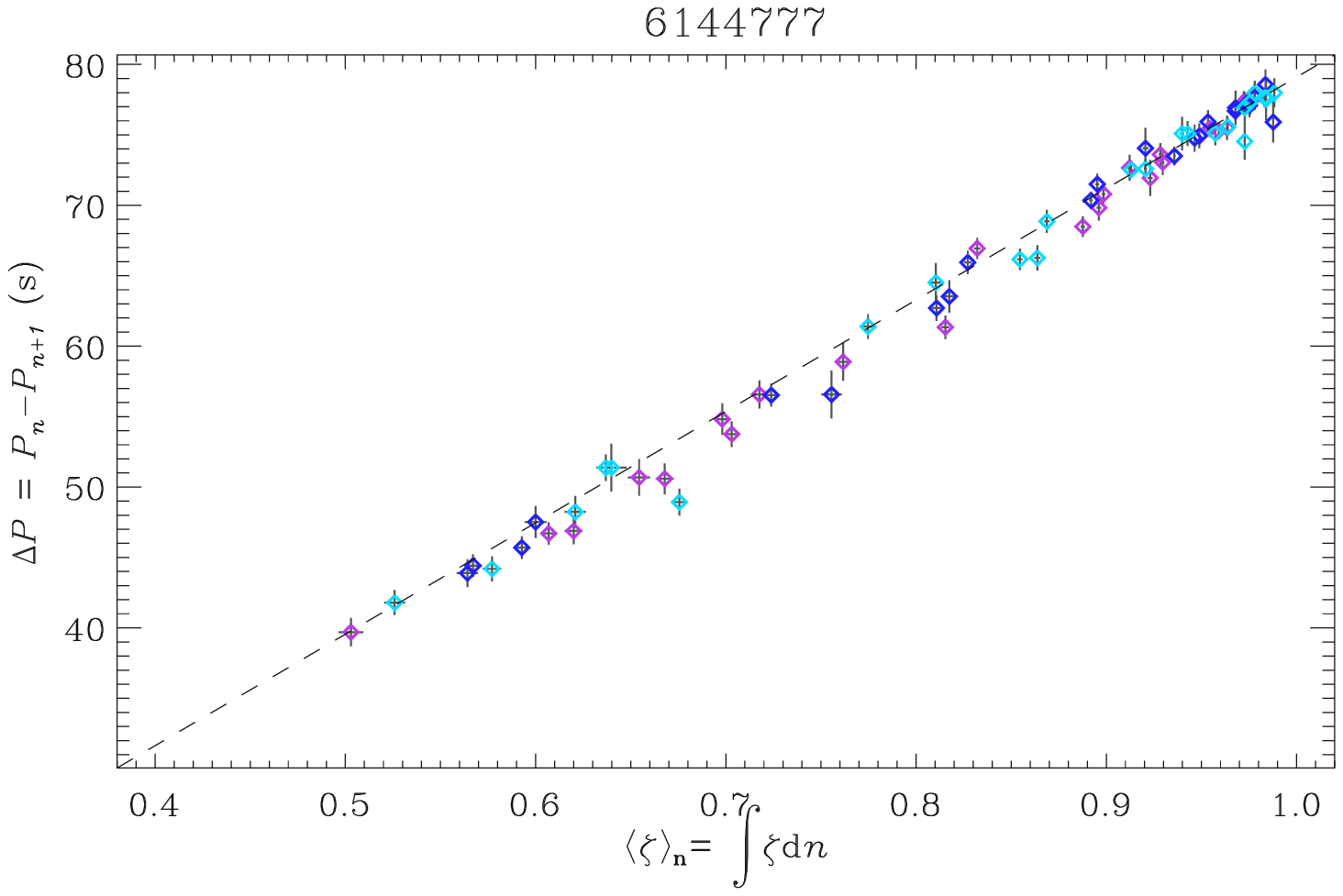}
\caption{Period spacings of the  RGB star KIC 6144777.
\emph{Top:} plot as a function of the arithmetical mean value $(\zeta(\nu_\nm) +
\zeta(\nu_{\nm+1}))/2.$
\emph{Bottom:} plot as a function of the mean value $\zmoyn$. The colors code the azimuthal orders, as in Fig.~\ref{fig-fit-6144777}; the dashed line indicates the 1:1 relation; 1-$\sigma$ uncertainties on both the spacings and the mean values of $\zeta$ are indicated by vertical and horizontal error bars.}
\label{fig-DP}
\end{figure}

\begin{figure}
\includegraphics[width=8.8cm]{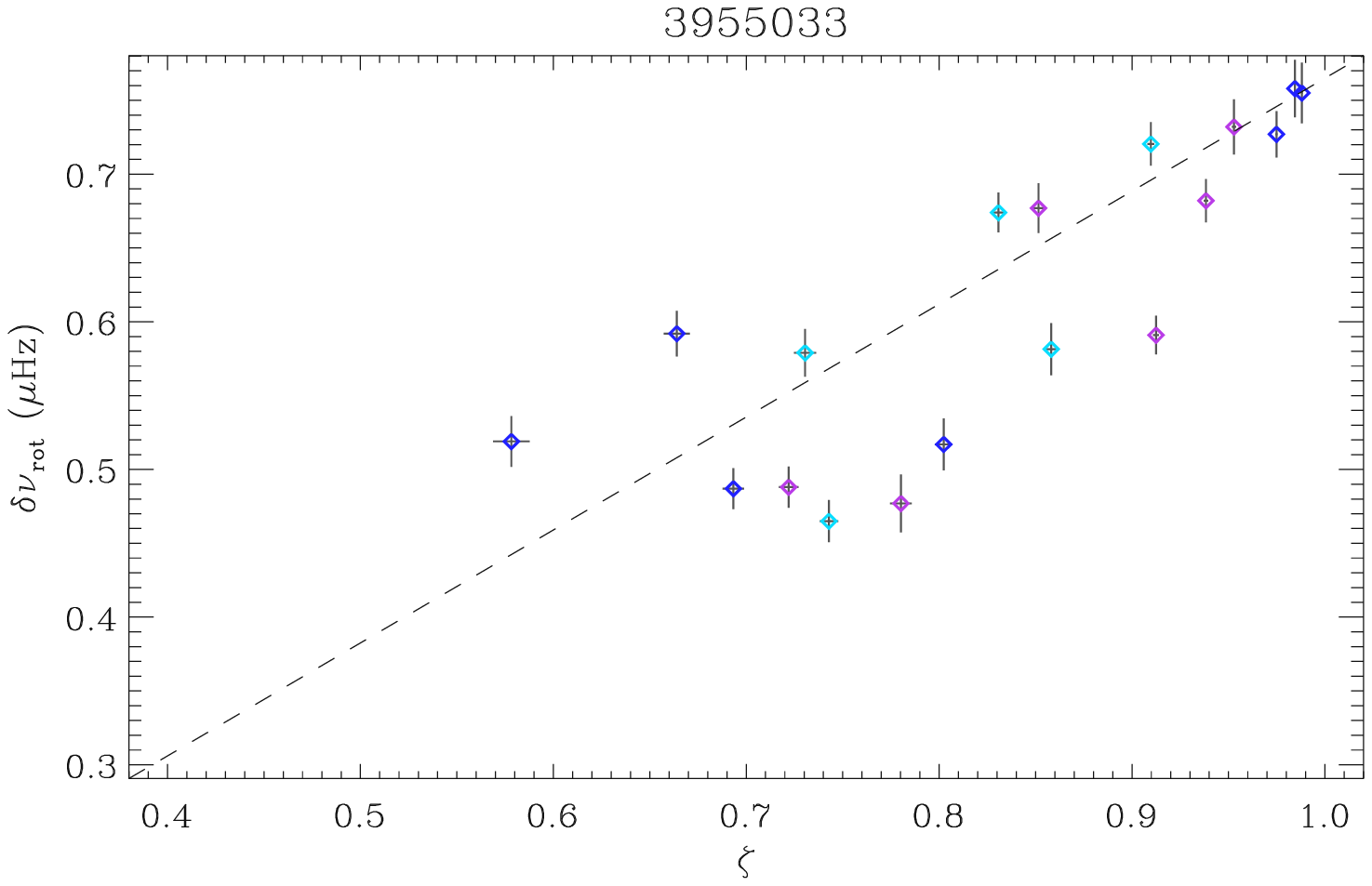}
\includegraphics[width=8.8cm]{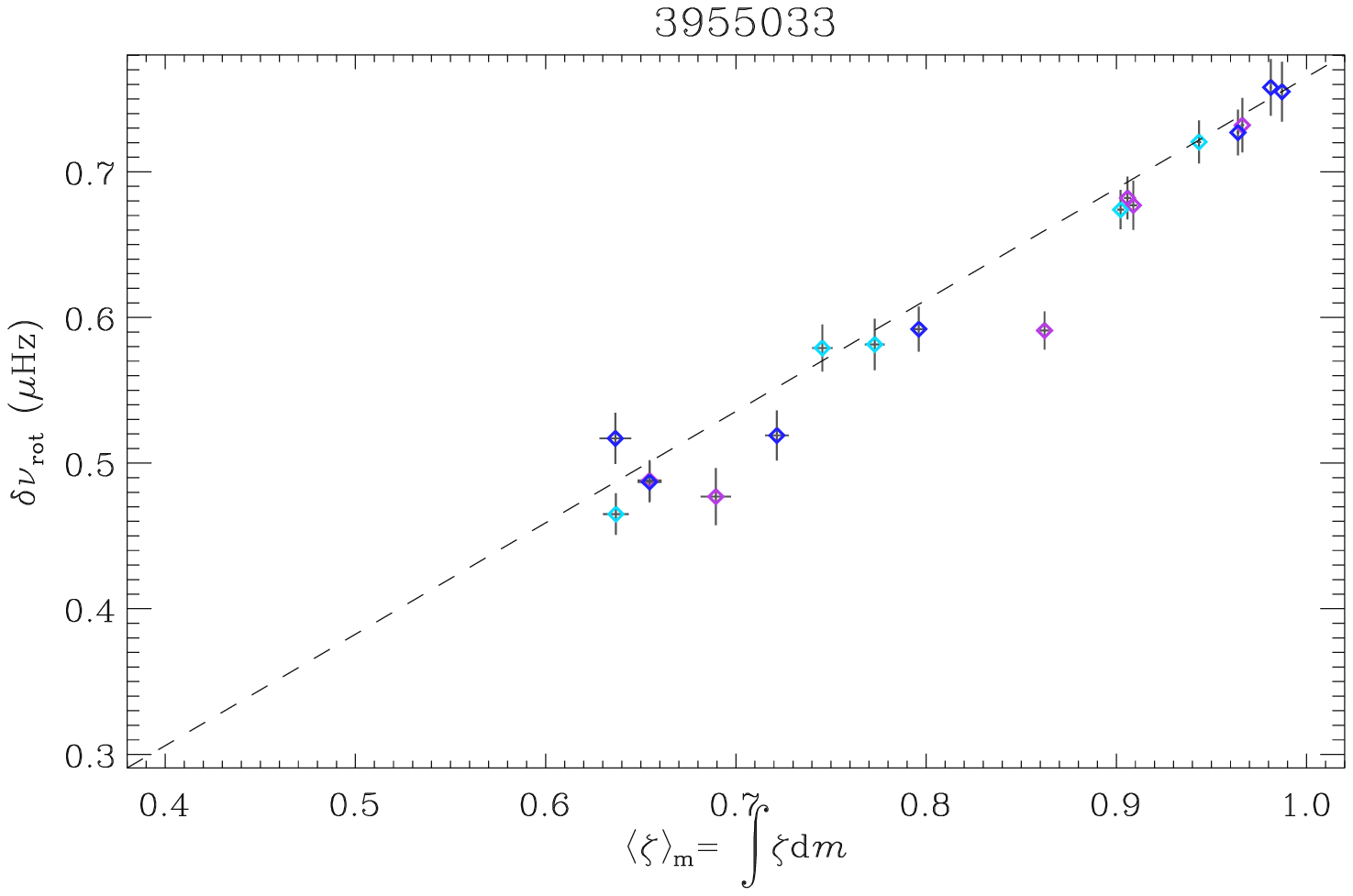}
\caption{Mean rotation splittings of the  RGB star KIC 3955033. \emph{Top:} plot as a function of $\zeta$.
\emph{Bottom:} plot as a function of the mean value $\zmoym$. The
colors code the azimuthal orders; the dashed line indicates the 1:1 relation; 1-$\sigma$ uncertainties on both the splittings and the mean values of $\zeta$ are indicated by vertical and horizontal error bars.} \label{fig-rotation}
\end{figure}

\subsection{Relationships with $\zeta$}

With the identification of the mixed-mode pattern, we aim to verify the relevance of the use of $\zmoyn$ for the period spacings, to test the relevance of $\zmoym$ for the rotational splittings, and further test the predictions for the mode widths and heights.

\subsubsection{Period spacings}

Period spacings were fitted with different functions of $\zeta$, according either to the integrated value $\zmoyn$ (Eq.~\ref{eqt-zeta-P}) or to the arithmetical mean $\zeta = (\zeta_\nm + \zeta_{\nm+1})/2$. The resulting plots are shown in  Fig.~\ref{fig-DP}. When $\zmoyn$ is not used, one remarks that the $\DP (\zeta)$ relation shows a modulation that results from the concavity of $\zeta$. When $\zeta$ is close to unity for gravity-dominated mixed modes, no modulation is seen; in the range [0.7,\,0.9], where the function is convex, the period spacings are larger than predicted; below 0.7, where the function is concave, the period spacings are smaller than expected. The relation between $\DP$ and $\zmoyn$ does not show such a modulation. Furthermore, the fit with $\zmoyn$ is nearly linear, with residuals two times smaller. From this comparison, we confirm that the use of $\zmoyn$ is preferable for fitting the period spacings.

\begin{figure}[t]
\includegraphics[width=8.8cm]{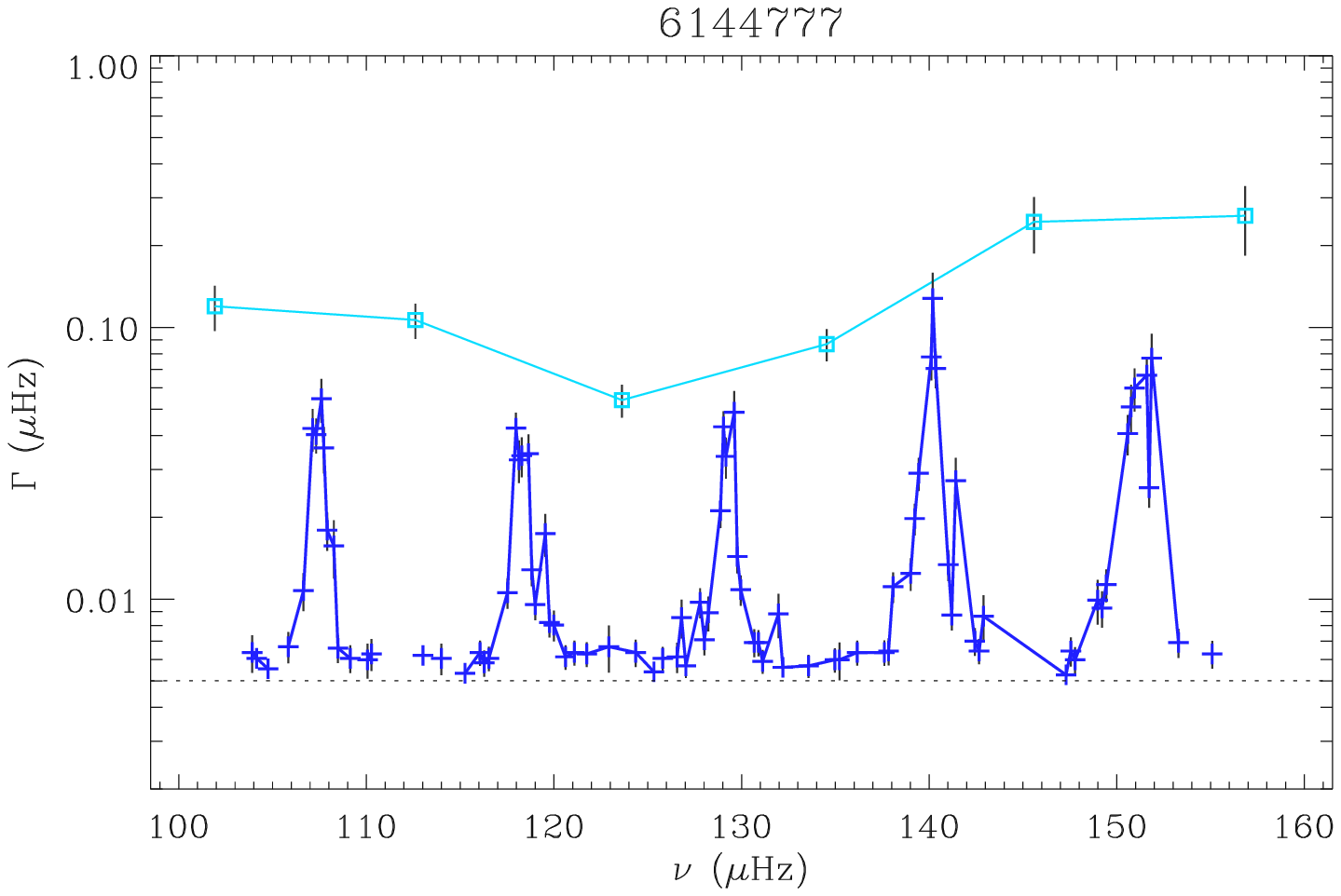}
\includegraphics[width=8.8cm]{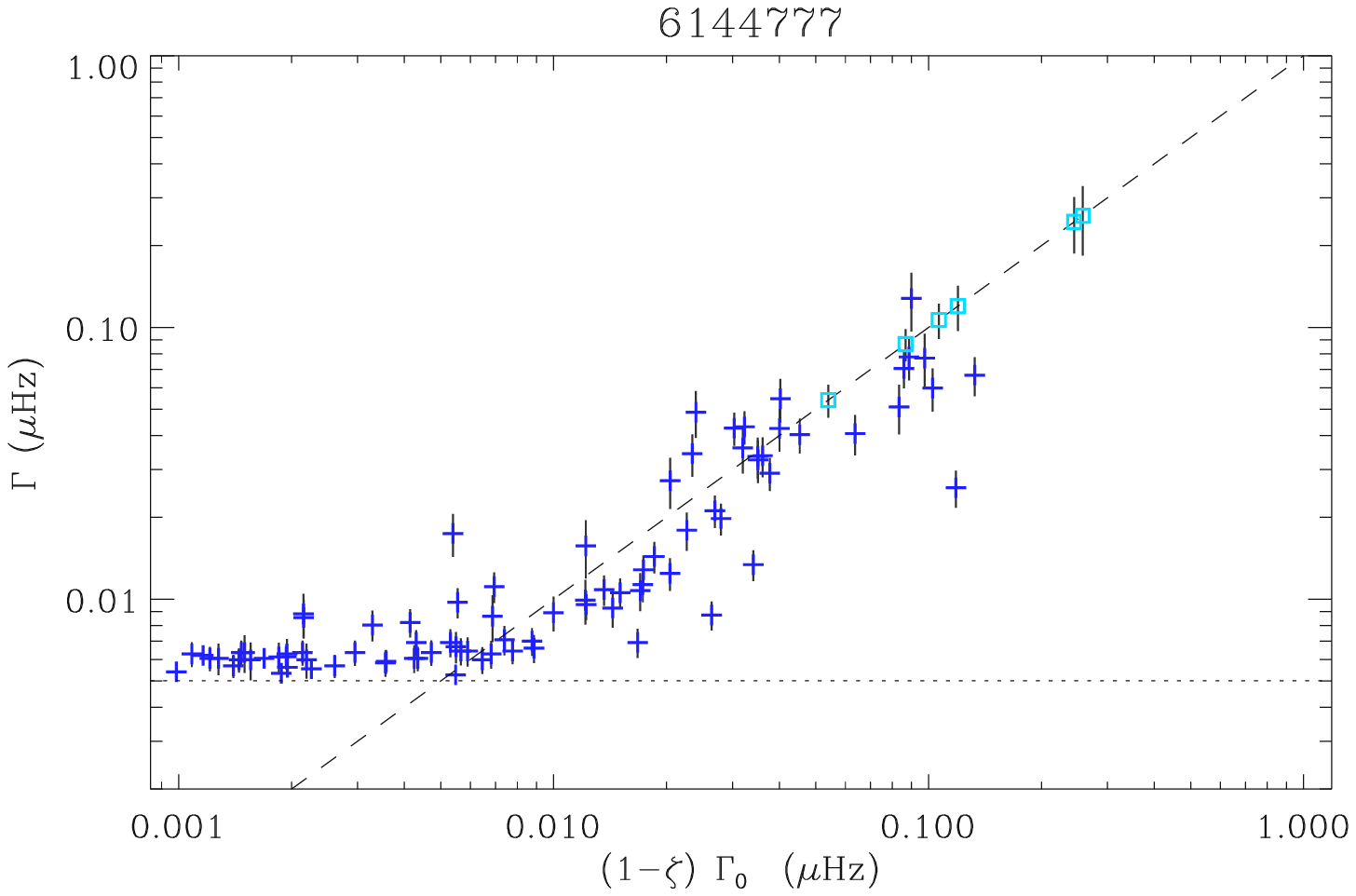}
\caption{Mode widths as a function of the cyclic frequency (\emph{top}) or as a function of  $(1-\zeta) \Gamma_0$ (\emph{bottom}), for the RGB star KIC 6144777. Radial modes are plotted with square symbols and dipole mixed modes with {\scriptsize{+}}; 1-$\sigma$ uncertainties on $\Gamma$ are indicated by vertical error bars. The value  $2\dfres/\pi$ (Eq.~\ref{eqt-zeta-H}) plotted as a dotted line is proportional to the 4-year long frequency resolution. In the bottom plot, radial
modes have been considered too, assuming they have $\zeta=0$ as
pure pressure modes.  The dashed line indicates the 1:1 relation.}\label{fig-gamma}
\end{figure}

\begin{figure}
\includegraphics[width=8.8cm]{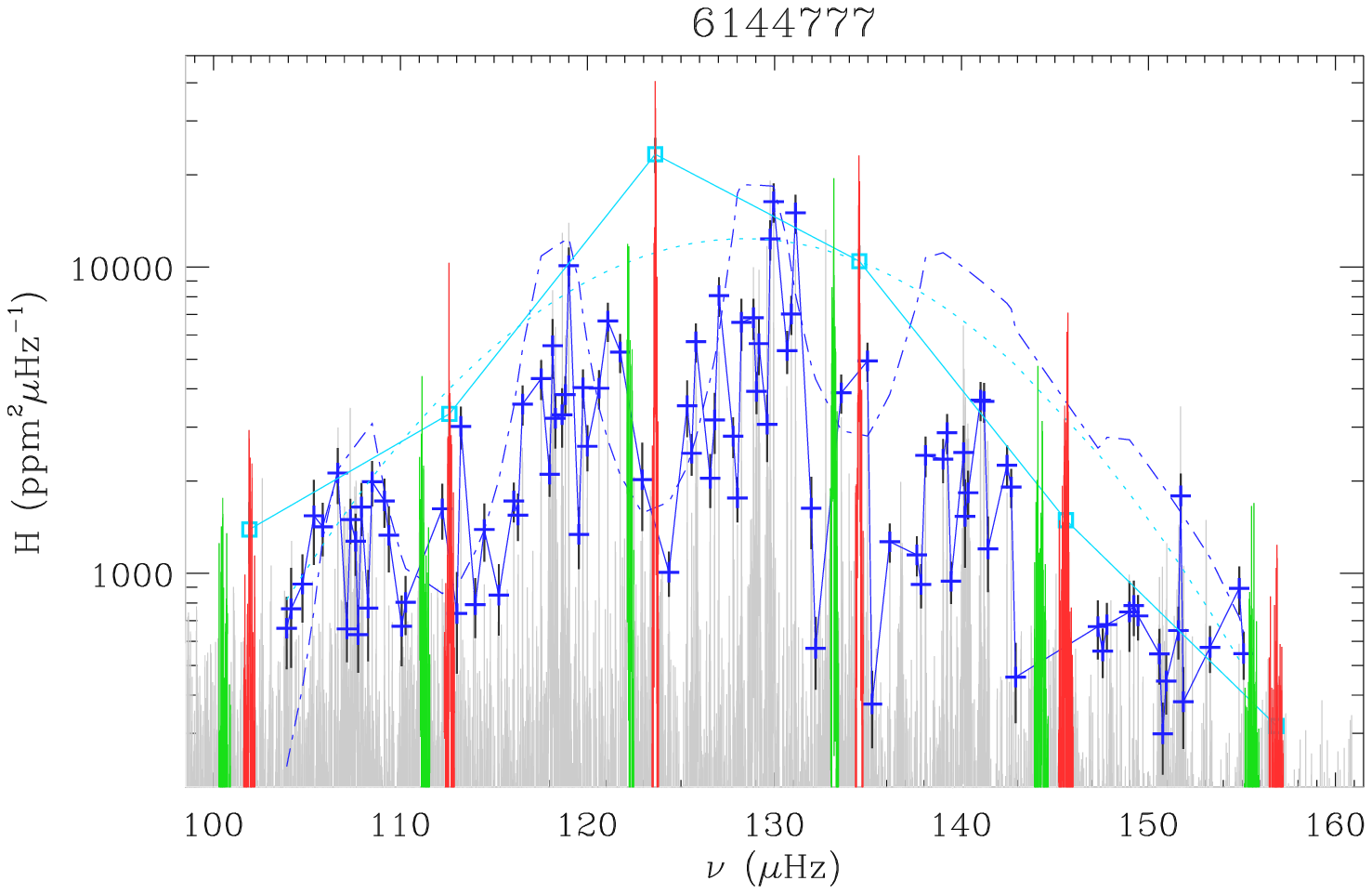}
\includegraphics[width=8.8cm]{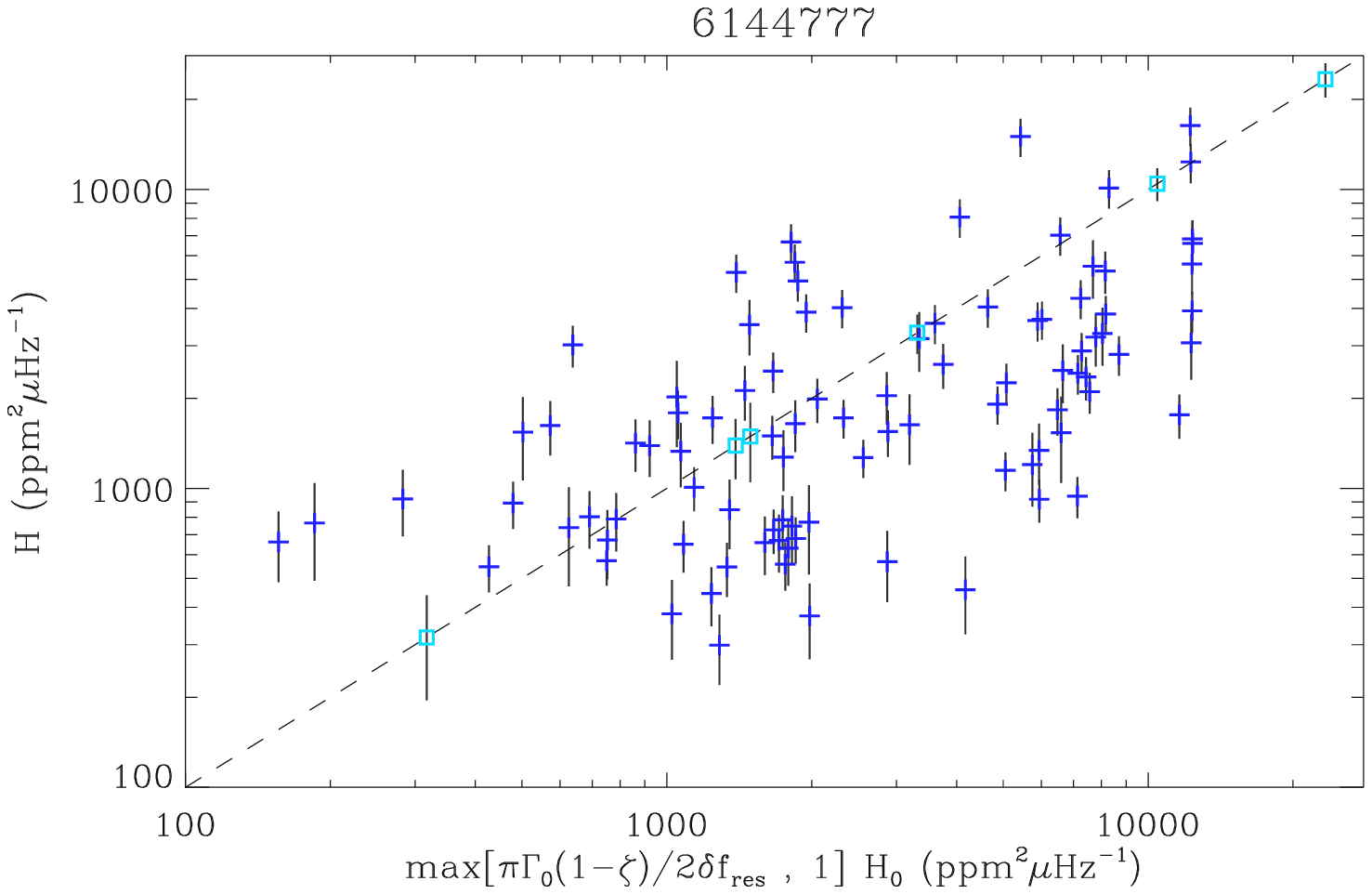}
\caption{Mode heights as a function of the cyclic frequency (\emph{top}) or as a function of the radial mode height, modified when the modes are not resolved, (\emph{bottom}), for the RGB star KIC 6144777. Radial modes are plotted with square symbols and dipole mixed modes with {\scriptsize{+}}. The Fourier spectrum is plotted in red (green) for emphasizing the radial (quadrupole) modes. The dot-dashed line provides the expected heights of dipole modes, under the assumption that the power excess mimics a Gaussian relation (dotted line).  The dashed line indicates the 1:1 relation. }\label{fig-power}
\end{figure}

\subsubsection{Rotational splittings}

We performed similar test for the rotational splittings. We a
priori excluded a dependence on $\zeta(\nu_{\nm,0})$, since we
clearly observe asymmetrical splittings (see below, Section \ref{asymmetry}) that cannot be
reproduced with $\zeta(\nu_{\nm,0})$. In fact, the rotation rate
of KIC 6144777 is not important enough to observe any difference
between the variations with either $\zeta$ or $\zmoym$. We
therefore performed the fit of the star KIC 3955033
(Fig.~\ref{fig-fit-3955033}), which shows a much more rapid rotation
(Fig.~\ref{fig-rotation}). From the comparison of $\dnurot(\zeta)$
and $\dnurot(\zmoym)$, we derive that this latter expression is
more convenient since it provides a $\chi^2$ ten times smaller than when using $\zeta$, associated with a much more precise estimate of the core rotation: $\dnurotcore =  765 \pm 10$\,nHz with $\zmoym$, versus $\dnurotcore =  730 \pm 50$\,nHz with $\zeta$. From this test, we conclude positively about the relevance of the use of $\zmoym$ for the rotational splittings.

\subsubsection{Widths, amplitudes, and heights}

As expected from Eq.~(\ref{eqt-largeur}), the mixed-mode width
shows large variations: pressure-dominated mixed modes have widths
comparable to those of the radial modes, contrary to
gravity-dominated modes that are much thinner (Fig.~\ref{fig-gamma}, top panel). Figure~\ref{fig-gamma} also shows the validity of Eq.~(\ref{eqt-largeur}), with the mixed-mode width
proportional to $(1-\zeta)$, except for low values where the
observations resolution hampers the measurement of very thin
widths. The precision of the fit is limited by the stochastic excitation, especially for long-lived peaks: the presence or absence of signal in a single frequency bin can modify the width in large proportion. This limit added to the limitation in frequency resolution does not allow us to test if small additional radiative damping affects the gravity-dominated mixed modes \citep{2009A&A...506...57D,2014A&A...572A..11G}.

As shown by \cite{2015A&A...584A..50M}, Eq.~(\ref{eqt-zeta-a2}) has a strong theoretical justification,
since it expresses the conservation of energy: the sum of all the energy distributed in the mixed modes corresponds to the energy expected in the single pure pressure mode that should exist in absence of any coupling. So, our result is in line with the findings of \cite{2012A&A...537A..30M}, who measured that, except for depressed modes, the observed total visibility of  dipole modes matches the theoretical expectations.

Due to the stochastic nature of the excitation, the mode heights show a large spread (Fig.~\ref{fig-power}). Dips in the distributions occur when modes are not resolved. It is however
clear in Fig.~\ref{fig-power} bottom that the dipole mode heights follow the radial distribution
according to the trend of Eq.~(\ref{eqt-zeta-H}). We note that all mixed modes associated with a given pressure radial order show a systematic behavior. For instance, all mixed modes of KIC 6144777 in the frequency range [137, 143\,$\mu$Hz] associated with the pressure mode $\np = 11$ show lower amplitudes than expected from the Gaussian fit of the power excess. Such a behavior recalls us that the excitation of a mixed mode is due to its acoustic component.

\subsection{Validation}

From these two case studies and from other examples shown in Appendix, we can conclude that the asymptotic fits are relevant at all evolutionary stages, when the signal-to-noise ratio is high enough. So, the equations developed in Section~\ref{parametres} allow us to depict the mixed-mode spectrum with a very high accuracy, when the integrated values $\zmoyn$ and $\zmoym$ are considered for the period spacings and the rotational splittings, respectively. Up to now, only red clump stars showing buoyancy glitches cannot be fitted with a single set of parameters.

\begin{figure}
\includegraphics[width=8.8cm]{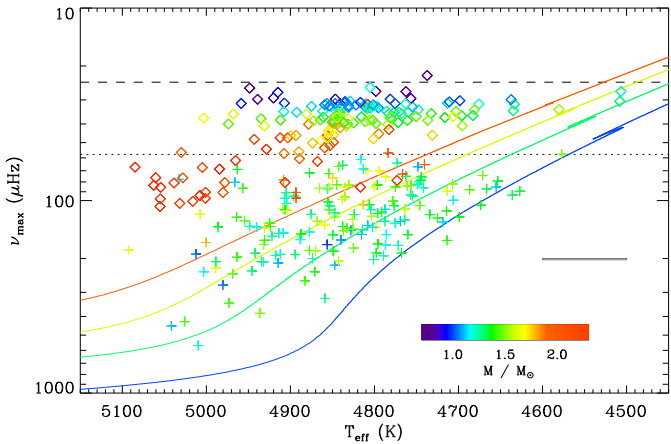}
\caption{Seismic diagram of the \nombre\ red giants studied in this work, with $\numax^{-1}$ used as a proxy for the luminosity. The color codes the stellar mass. Stars on the RGB are plotted with $+$ symbols, red clump stars with $\diamond$. The dotted and dashed lines indicate the limit of the visibility of mixed modes for RGB and clump stars, respectively, as defined by Eq.~(\ref{eqt-limite-visi-gm}). Evolution tracks on the RGB, computed with MESA for solar metallicity \citep{2018arXiv180204558G}, are shown for the stellar masses 1.0, 1.3, 1.6, and 1.9\,$M_\odot$. The error box indicates the typical 1-$\sigma$ uncertainties on $\Teff$ and $\numax$.}
\label{fig-HR}
\end{figure}

\section{Asymptotic period spacings and gravity offsets\label{application-asymptotic}}

In this section, we show how previous findings can be used to derive accurate period spacings. We also explore the variation of the gravity offsets $\epsg$ with stellar evolution. These studies rely on the determination of the pure-gravity mode pattern.

\begin{figure}
\begin{center}
\includegraphics[width=5.6cm]{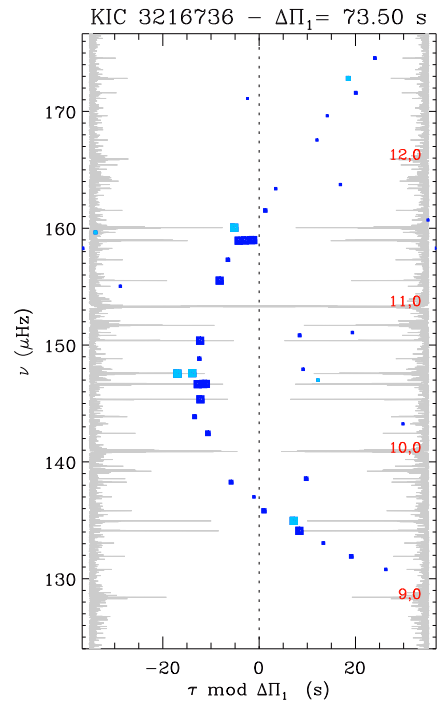}
\end{center}
\caption{Stretched-period \'echelle diagram of KIC 3216736, the only RGB star in our sample showing buoyancy glitches. The spectrum is simple, since only $m=0$ dipole mixed modes are present, but shows a large-period modulation instead of the expected vertical alignment. Modes plotted in light blue are pressure dominated; extra peaks that do not follow the global trend are either $\ell=3$ modes or $\ell=2$ mixed modes.  Red figures indicate the radial orders of the radial modes. For clarity, the power spectrum density is also plotted twice, top to tail.
}\label{fig-ech-3216736}
\end{figure}

\begin{figure}
\includegraphics[width=8.8cm]{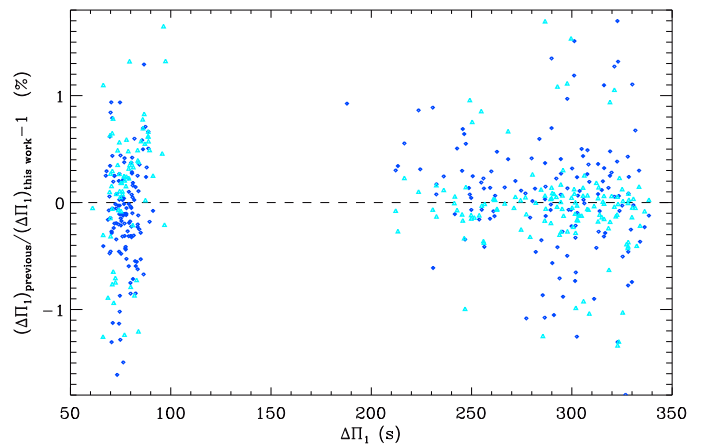}
\caption{Comparison of the asymptotic period spacings with previous values. Light blue triangles show the bias in period spacings computed under the assumption $\epsg=0$ \citep{2014A&A...572L...5M}, whereas dark blue diamonds are free of this hypothesis \citep{2016A&A...588A..87V}.}\label{fig-comparDPi}
\end{figure}

\begin{figure}
\includegraphics[width=8.8cm]{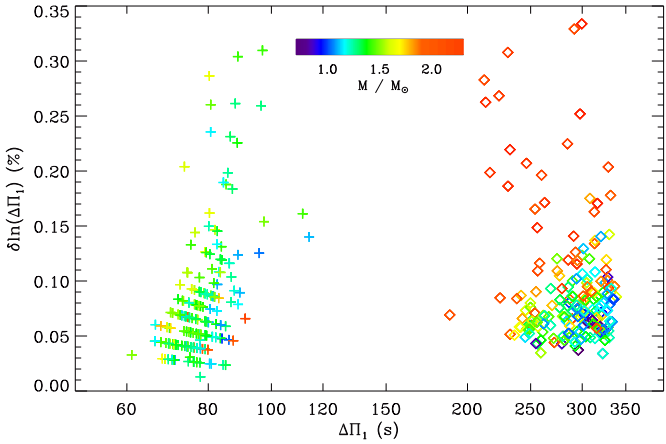}
\caption{Relative precision of the asymptotic period spacings. Same style as in Fig.~\ref{fig-HR}.}\label{fig-accuracy-Tg}
\end{figure}

\begin{figure*}
\includegraphics[width=15cm]{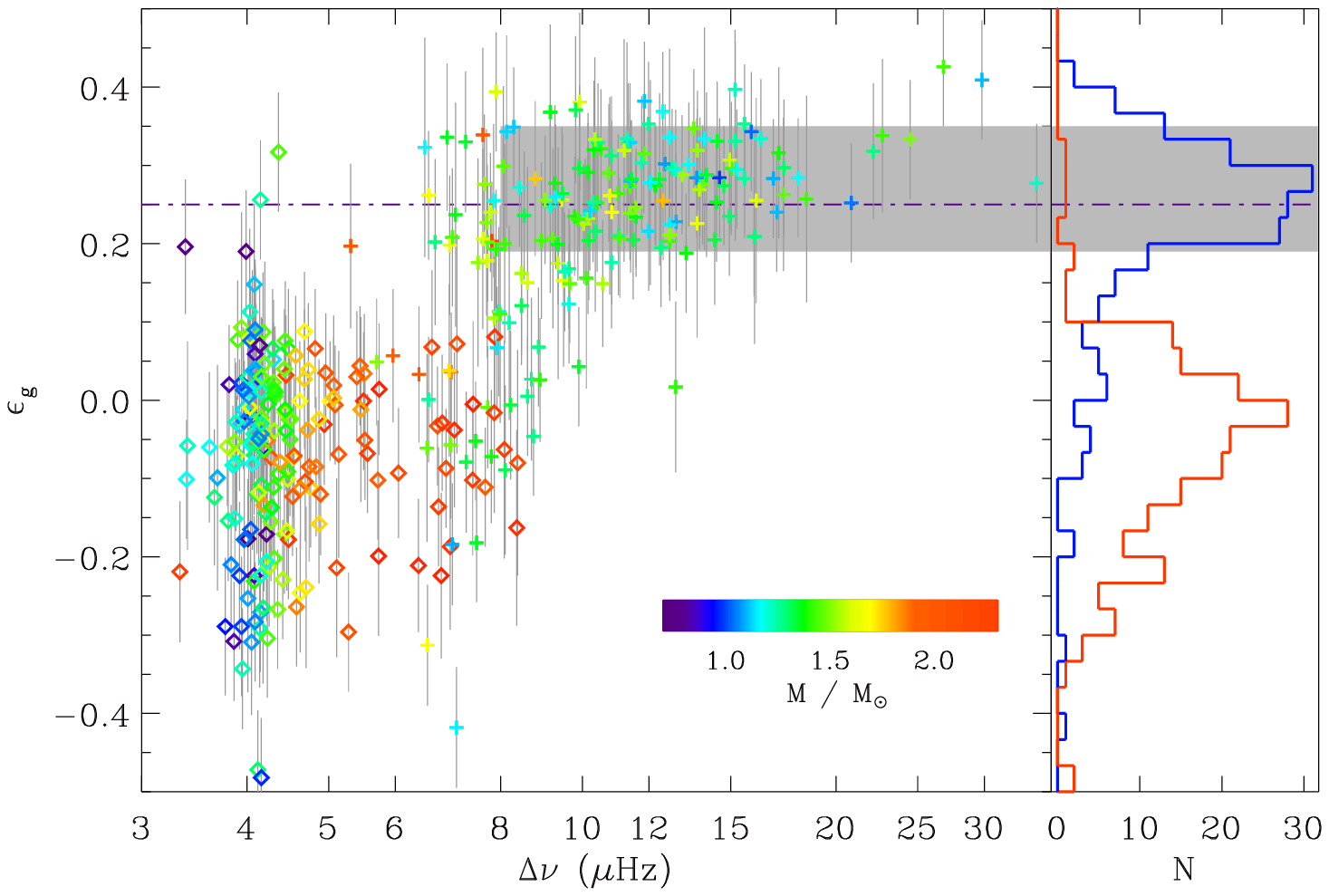}
\caption{\emph{Left:} variation of $\epsg$ with $\Dnu$, with the same style as Fig.~\ref{fig-accuracy-Tg}. The horizontal dark gray domain corresponds to the expected range predicted for RGB stars by \cite{2016PASJ...68..109T}, whereas the dot-dashed line shows the value $\epsg{}\ind{,as} =1/4$ derived from the asymptotic expansion \citep{1986A&A...165..218P}.  Uncertainties on $\epsg$ are indicated by vertical lines; uncertainties on $\Dnu$ are smaller than the symbol size. 
\emph{Right:} histograms of the distributions of $\epsg$ on the RGB (blue curve) and in the red clump (red curve). The dot-dashed line and the gray domain have the same meaning as indicated above.}
\label{fig-epsg}
\end{figure*}

\begin{table*}
\caption{Period spacings and gravity offsets}\label{tab-epsg}
\begin{tabular}{rrrrrrr}
\hline
 KIC     & $\numax\qquad$      & $\Dnu\qquad$       & $\Tg\qquad$ & $q\qquad$  & $\epsg\qquad$  & $\dnurot\quad$\\
         & ($\mu$Hz$)\qquad$   & ($\mu$Hz$)\qquad$  & (s$)\qquad$ &            &                & (nHz$)\quad$  \\
  \hline
 1576469 &$ 90.60\pm 0.98$&$ 7.41\pm0.04$&$ 284.80\pm0.64$&$0.23\pm0.03$&$-0.102\pm0.096$&$  67\pm  6$\\
 1723700 &$ 39.42\pm 0.57$&$ 4.48\pm0.04$&$ 323.40\pm0.17$&$0.24\pm0.04$&$ 0.066\pm0.043$&$  57\pm  5$\\
 2437976 &$ 89.37\pm 1.10$&$ 8.22\pm0.05$&$  74.70\pm1.00$&$0.10\pm0.02$&$-0.006\pm0.095$&$ 320\pm 30$\\
 2443903 &$ 66.76\pm 0.90$&$ 7.01\pm0.04$&$  71.10\pm0.02$&$0.12\pm0.02$&$-0.184\pm0.078$&$ 360\pm  4$\\
 3955033 &$106.10\pm 1.24$&$ 9.23\pm0.05$&$  74.65\pm0.06$&$0.13\pm0.02$&$ 0.207\pm0.115$&$ 765\pm 10$\\
 5024476 &$ 68.66\pm 0.75$&$ 5.73\pm0.04$&$ 299.60\pm1.00$&$0.27\pm0.03$&$-0.199\pm0.102$&$  63\pm  6$\\
 5112373 &$ 43.82\pm 0.59$&$ 4.63\pm0.04$&$ 240.30\pm0.14$&$0.19\pm0.02$&$-0.246\pm0.058$&$  37\pm  3$\\
 6144777 &$128.23\pm 1.50$&$11.03\pm0.05$&$  79.05\pm0.04$&$0.13\pm0.02$&$ 0.210\pm0.055$&$ 244\pm  5$\\
10272858 &$341.45\pm 6.16$&$22.71\pm0.14$&$  96.90\pm0.30$&$0.19\pm0.02$&$ 0.338\pm0.098$&$ 660\pm 20$\\
11353313 &$127.29\pm 1.46$&$10.75\pm0.05$&$  76.95\pm0.06$&$0.14\pm0.02$&$ 0.290\pm0.088$&$ 465\pm  7$\\
  \hline
\end{tabular}

The list of the full data set with \nombre\ red giants showing an uncertainty in $\epsg$ less than 0.15 is available on line as a CDS/VizieR document.
\end{table*}

\subsection{Observations}

Our analysis was conducted over \nombre\ red giants at various evolutionary stages, mainly from \cite{2014A&A...572L...5M} and \cite{2016A&A...588A..87V}, with stars also considered in \cite{2012Natur.481...55B}, \cite{2012A&A...541A..51K}, \cite{2014A&A...564A..27D}, and \cite{2015A&A...579A..83C}.
Data were obtained as for the two stars considered in Section~\ref{casestudy}. When available, effective temperatures are from APOGEE spectra \citep{2017ApJS..233...25A}. Selection criteria are mainly based upon the noise level, with \Kepler\ magnitudes brighter than 12 on the low RGB or 14 for more evolved stars. Following the method exposed in Section \ref{identi_mm}, we need data with a S/R high enough to allow the identification of gravity-dominated mixed modes. When such modes are too few, measurements are impossible. This condition induces a selection bias, specifically addressed in Section \ref{application-observability}.

The $\nombre$ stars that were analyzed are shown in a seismic diagram (Fig.~\ref{fig-HR}). We considered stars from the low RGB (Fig.~\ref{fig-fit-10272858}) to more evolved RGB stars (Fig.~\ref{fig-fit-11353313}). The spectrum of the evolved RGB star KIC 2443903 (Fig.~\ref{fig-fit-2443903}) corresponds to a case near the limit of visibility of gravity-dominated mixed modes, with a mode density $\dens \simeq 22.4$ close to the limit value above which the detection is impossible (Section \ref{application-observability}). The fitting process for red clump stars can be achieved only when the amplitude of the buoyancy glitches remains limited (Fig.~\ref{fig-fit-1723700}); the same limitation appears in the secondary red clump (Fig.~\ref{fig-fit-1725190}). In fact, except for red-clump stars with large buoyancy glitches \citep{2015ApJ...805..127C,2015A&A...584A..50M}, the asymptotic expansion provides a relevant fit. We could then obtain precise measurements of the asymptotic gravity parameters in Eq.~(\ref{eqt-asymp-g}) and of their uncertainties for a large number of stars. We must report one exception: KIC 3216736 is the only RGB star of our sample where we found buoyancy glitches and could not provide a relevant fit of the spectrum, but only an \'echelle diagram based on stretched periods (Fig.~\ref{fig-ech-3216736}). Since we have tested more than 160 stars on the RGB, with a systematic approach, we can conclude that the most common case on the RGB is the absence of buoyancy glitches, as expected theoretically \citep{2015ApJ...805..127C}.

Characterizing the sample we studied in terms of bias is difficult. Apart from the RGB stars that were already studied in detail in previous works, we have mostly treated the stars with increasing KIC numbers. This systematic method implies that we did not introduce any further bias compared to the \Kepler\ sample of red giants. Considering a high enough S/R, which is almost equivalent to select bright stars in the red giant domain, is not supposed to introduce biases either. Contrary to many previous studies, we are not limited to stars showing rotational splittings smaller than the confusion limit ($\dnurot \le \numax^2 \Tg$). However, the presence of a strong cutoff (Section \ref{application-observability}) limits the sample, when gravity-dominated mixed modes disappear. Red-clump stars with strong buoyancy glitches are absent in our data set since the fitting process requires then to account for the extra-modulation, which can be quite large (about $\Tg / 10$). When mixed modes are depressed, the low height-to-background ratio of the mixed modes allows the measurement of $\Tg$ \citep{2017A&A...598A..62M} but is not enough for fitting the pattern. Both cases  deserve specific care beyond the scope of this work.

\subsection{Pure gravity modes}

The identification of the mixed modes depicted in Section \ref{identi_mm} allows us to retrieve the periods of the pure gravity modes and to infer global asymptotic parameters of the gravity components. We compute these periods from the mixed-mode frequencies $\nu$, using Eqs.~(\ref{eqt-asymp}) and (\ref{eqt-g}),
\begin{equation}\label{eqt-pur-g}
  {1\over \nug} = {1\over \nu} - {\Tg\over\pi} \ \hbox{atan}\left( {\tan \thetap\over q } \right)
  .
\end{equation}
Close to each radial mode, when $\thetap$ varies from values less than but close to $\pi/2$ to values higher than but close to $-\pi/2$, the atan correcting term introduces an offset of $-\Tg$, which in fact allows to relate the $(\dens +1)$ mixed modes in the $\Dnu$-wide interval to $\dens$ only gravity modes. In order to use all mixed modes, including the $|m|=1$ components, we corrected first the rotational splittings, using Eq.~(\ref{eqt-zeta-rotation2}) in order to obtain $\nu$ values that are corrected from the rotational splitting.

From the periods of the gravity modes $1/\nug$, we could then derive the asymptotic parameters $\Tg$ and $\epsg$, assuming the first-order asymptotic expression for pure gravity modes (Eq.~\ref{eqt-asymp-g}). In practice, a first estimate of $\Tg$ derived from the formalism of \cite{2015A&A...584A..50M} and \cite{2016A&A...588A..87V} is used in Eq.~(\ref{eqt-pur-g}), then iterated with a least-square fit of the linear variation of the gravity modes (Eq.~\ref{eqt-asymp-g}).

\subsection{Asymptotic period spacings}

Up to now, measurements of $\Tg$ considering that $\epsg$ is a free parameter have been obtained for a few stars only \citep[][for 3 and 22 observed stars, respectively]{2016A&A...588A..82B,2018A&A...610A..80H}. The offset $\epsg$ being arbitrarily fixed, \cite{2012A&A...540A.143M} and \cite{2014A&A...572L...5M} reported a very high precision for the period spacings, of typically 0.1\,s for stars on the RGB and 0.3\,s in the red clump. However, owing to the choice of $\epsg=0$, their period spacings were slightly affected by a bias of about a fraction of $\numax \Tg^2$. The values reported by \cite{2016A&A...588A..87V}, free of any hypothesis on $\epsg$, are not biased but show uncertainties typically five to fifteen times higher than the new values.  Their comparison with our data confirms the absence of systematic offsets (Fig.~\ref{fig-comparDPi}). So, the new method ensures accuracy, in the sense that the measurements of $\Tg$ are now free of any hypothesis on $\epsg$ and prove that the asymptotic expansion for gravity modes (Eq.~\ref{eqt-asymp-g}) is relevant. The relative accuracy we obtained for the period spacings, assuming Eq.~(\ref{eqt-asymp-g}), is shown in Fig.~\ref{fig-accuracy-Tg}. The median relative accuracies on the RGB and in the red clump are similar, of about $7\ 10^{-4}$. They translate, respectively, into 0.06\,s on the RGB and 0.22\,s in the red clump; \cite{2018A&A...610A..80H} reach a similar precision.

\subsection{Gravity offsets $\epsg$\label{gravity-offsets}}

We could measure $\epsg$ for a large set of stars. We however have to face the indetermination of $\epsg$ modulo 1:  we simply assume that $\epsg$ is in the range $[-0.5,\,0.5]$. The $\epsg$ values computed for the set of stars presented in the paper is given in Table~\ref{tab-epsg} and plotted in Fig.~\ref{fig-epsg}, where an histogram is also given. The complete table is given online only. Uncertainties on $\epsg$ are small, related  to the uncertainties in $\Tg$ by
\begin{equation}\label{eqt-asymp-depsg}
  \delta\epsg = {\delta\Tg \over \numax \ \Tg^2}
  .
\end{equation}
This relation comes from the derivative of Eq.~(\ref{eqt-asymp-g}). As a result, the median uncertainties are of about 0.08 on the RGB and 0.06 in the red clump.

We noticed that the median value of $\epsg$ on the RGB  is in fact close to 1/4, which is the expected asymptotic value in absence of stratification below the convection zone \citep{1986A&A...165..218P}, derived from the contribution $\ell/2 - \varepsilon\ind{as}$ with $\ell=1$ and $\varepsilon\ind{as}=1/4$. Hence, we inferred that the degeneracy on the determination of $\epsg$ is removed. We then noted a slight decrease in $\epsg$ when stars evolve on the RGB, with an accumulation of values close to 0 for red-clump stars. \cite{2018A&A...610A..80H} reported values of  $\epsg$  in the range $[-0.2,\,0.5]$ for 21 stars on the RGB, but did not identify the accumulation of values in the range $[0.20,\,0.35]$ predicted by  \cite{2016PASJ...68..109T} for stars on the low RGB. Our measurements fully confirm this prediction. From a check of their data set, we interpret the differences in $\epsg$ as resulting from less precise gravity spacings when large rotational splittings apparently modify the period spacings. As made clear by the recent theoretical developments of the asymptotic expansion \citep{2016PASJ...68...91T,2016PASJ...68..109T}, the accurate measurement of the leading-order term $\Tg$ is necessary to provide reliable estimates of $\epsg$.

We can study the variation of $\epsg$ along stellar evolution. On the RGB, the asymptotic expansion predicts $\epsg=1/4-\vartheta$ \citep{1986A&A...165..218P}, where $\vartheta$ is a measure of the stratification just below the convection zone. From this dependence, we can infer that the term $\vartheta$ is certainly very small for most stars on the low RGB. Higher values are suspected for evolved RGB stars, but with too few stars to firmly conclude, whereas lower values are seen for evolved RGB. We checked that the change of regime of $\epsg$ is not associated with the luminosity bump since it occurs for more evolved stars than our sample \citep{2018ApJ...859..156K}. In the red clump, the $\vartheta$ correction seems important, on the order of 0.3, with a larger spread than observed on the RGB.

An extended study of $\epsg$ can now be performed to use this parameter as a probe of the stratification occurring in the radiative region. This study is however beyond the scope of this work.


\begin{figure}[t]
\includegraphics[width=8.8cm]{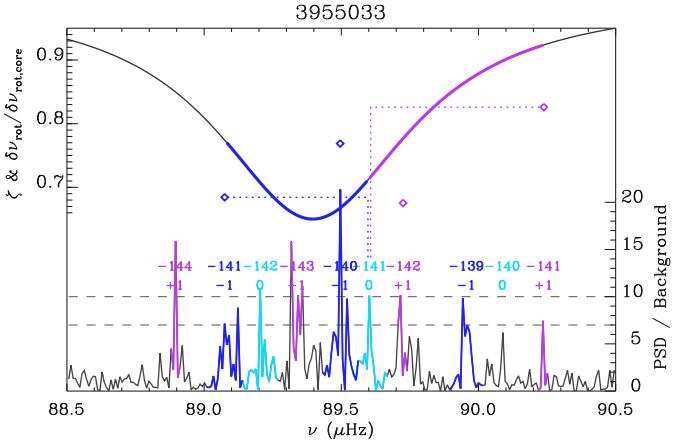}
\caption{Splitting asymmetry at low frequency in KIC 3955033. Each
dipole mixed mode of the spectrum is labelled with its radial and azimuthal orders. The rotational splittings of the radial orders from $-142$ to $-140$, plotted with diamonds, do no match the function $\zeta$. Only the multiplet with $\nm=-141$ is complete: the $m=+1$ splitting is much larger than the $m=-1$ splitting; the colored regions indicate the ranges over which the function $\zeta$ is integrated for the components of the multiplet $\nm=-141$. The dashed lines indicate height-to-background values of 7 and 10. }
\label{fig-asy-3955033}
\end{figure}

\section{Rotation\label{application-rotation}}

The fits based on the function $\zeta$ also allow us to analyze rotational splittings in detail.

\subsection{Splitting asymmetry\label{asymmetry}}

Recently, asymmetries in the rotational splittings were reported by \cite{2017A&A...605A..75D}, as the signature of the combined effects of rotation and mode mixing. Using both perturbative and non-perturbative approaches, they computed near-degeneracy effects and could fit the data. In fact, the asymptotic development of mixed modes also describes the combined effects of rotation and mode mixing, so that the rotational splittings based on $\zmoym$ (Eqs.~\ref{eqt-zeta-rotation2} and \ref{eqt-zmoy-m}) are not symmetric. Inversely, the symmetrical rotational splitting based on $\zeta$ (Eq.~\ref{eqt-zeta-rotation}) does not reproduce the observed asymmetry. Hence, observing asymmetrical triplets is a way to prove the relevance of the use of $\zmoym$ instead of $\zeta$.

Observing the asymmetry is challenging but possible for stars with a rapid rotation rate. As explained by \cite{2017EPJWC.16004005G,2018arXiv180204558G}, rapid rotation means $\dnurot \ge \Dnu / \dens$ for seismology. This rotation is however very slow in terms of interior structure, so that the formalism developed by \cite{2013A&A...549A..75G} and \cite{2014A&A...564A..27D}, summarized by Eq.~(\ref{eqt-zeta-rotation2}), remains relevant. It simplifies the study, as shown by \cite{2013A&A...554A..80O} who treated the case where rotational splittings can be as large as $\Dnu$. We fitted the mixed-mode spectrum of KIC 3955033 with both the symmetrical and asymmetrical splitting. At high radial order $\np$, it is hard to distinguish them. At low orders, when the rotational splittings exceed the mixed-mode spacings, the symmetrical splittings fail whereas the asymmetrical one provides a consistent solution along the whole spectrum. The radial order $\np=8$ is shown in Fig.~\ref{fig-asy-3955033}, the whole spectrum is shown in Fig.~\ref{fig-fit-3955033}.

\subsection{Surface rotation}

For stars on the low RGB, surface rotation can be inferred from the rotational splittings (Eq.~\ref{eqt-zeta-rotation}). The measurement is however difficult, since it results from an extrapolation at $\zeta=0$, when values are mostly obtained above $\zeta=0.6$ only (Fig.~\ref{fig-rotation}). The highest level of precision, hence the use of $\zmoym$ instead of $\zeta$, is required for deriving a correct estimate of the surface rotation. The case of KIC 3955033 is illustrative, with a negative surface rotation when using $\zeta$; the use of $\zmoym$ provides a null value ($5\pm20$\,nHz). This case also confirms the general situation shown by previous works  \citep{2013A&A...549A..75G,2016ApJ...817...65D,2017A&A...602A..62T}: deriving surface rotation can be achieved for the low RGB only.

\subsection{Stellar inclination}

From its ability to fit the gravity-dominated modes that carry useful information, the asymptotic fit can be used to derive the stellar inclination $i$. The amplitude of the $m=0$ component of the dipole multiplet is proportional to $\sin^2 i$ whereas the sum
of the amplitudes of the $m=\pm 1$ mode is proportional to $\cos^2 i$. From Eq.~(\ref{eqt-zeta-a2}), a correction factor of $1/(1-\zeta)$ should be applied on the amplitudes: its differential effect is however much below the precision one can get on $i$.

We tested our results on a set of stars for which the inclinations measured with other methods have been obtained. We checked that our results are relevant, with a precision limited by the uncertainties on the amplitude measurements. In order to avoid bias, we consider only peaks with a height-to-background ratio larger than 8. Nevertheless, we noted that the stochastic excitation of the modes induces a small bias for large inclinations. Equator-on inclinations, near $90^\circ$, cannot be retrieved precisely, with measurements reduced toward the range $70$--$80^\circ$. As a consequence, they are rare in our analysis. However, many stars show inclinations that, according to the uncertainties, are compatible with equator-on measurement, so that the bias does not affect the following analysis.

\begin{figure}
\includegraphics[width=8.8cm]{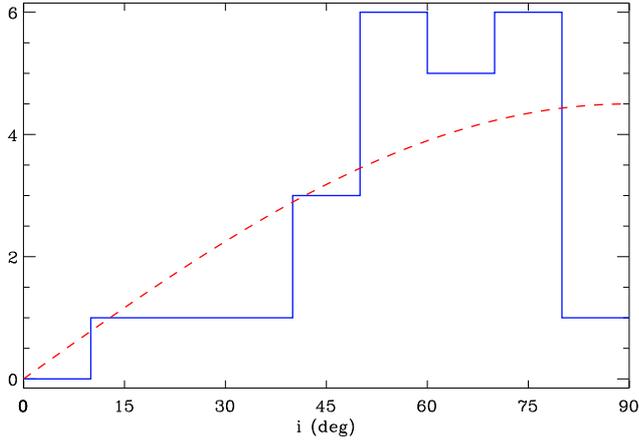}
\caption{Histogram of the inclinations measured in NGC 6819. The dashed line indicates the $\sin i$ distribution.
}\label{fig-i-histo}
\end{figure}

\begin{table}
\caption{Asymptotic and rotational parameters in NGC 6819}\label{tab-NGC6819-incli}
\begin{tabular}{lrrrrr}
  \hline
 KIC ID &$\Dnu$& $\Tg$ & $q$ & $\dnurot$&  $i$  \\
        &($\mu$Hz) &(s)  & & (nHz)& ($^\circ$) \\
  \hline
 4937056 &  4.76 &  291.0 & 0.21 &  90 & 60$\pm$15 \\
 4937257$^a$ &  4.13 &  292.1 & 0.19 &  27 & 72$\pm$13 \\
 4937770$^b$ &  7.82 &  161.0 & 0.18 &   $\times$ &  $\times$ \\
 4937775$^a$ &  7.33 &  226.3 & 0.21 & 110 & 75$\pm$15 \\
 5023953 &  4.74 &  293.9 & 0.24 &  50 & 51$\pm$28 \\
 5024327 &  4.72 &  269.5 & 0.20 &  55 & 56$\pm$13 \\
 5024404 &  4.78 &  242.6 & 0.25 & 110 & 80$\pm$10 \\
 5024414 &  6.47 &  283.0 & 0.30 &  90 & 45$\pm$20 \\
 5024476 &  5.73 &  299.5 & 0.24 &  56 & 71$\pm$11 \\
 5024582 &  4.76 &  323.5 & 0.22 &  70 & 55$\pm$18 \\
 5111718 & 10.59 &   88.4 & 0.12 & 410 & 69$\pm$21 \\
 5111949 &  4.81 &  319.0 & 0.28 &  35 & 66$\pm$15 \\
 5112072 & 10.08 &   91.9 & 0.15 & 350 & 72$\pm$12 \\
 5112361$^c$ &  6.19 &   99.0 & 0.12 & 350 & 70$\pm$20 \\
 5112373 &  4.63 &  240.2 & 0.19 &  37 & 47$\pm$18 \\
 5112387 &  4.70 &  267.2 & 0.28 &  84 & 25$\pm$17 \\
 5112401 &  4.03 &  311.0 & 0.26 &  50 & 54$\pm$13 \\
 5112467 &  4.75 &  285.2 & 0.25 &  90 & 61$\pm$12 \\
 5112491 &  4.68 &  324.3 & 0.30 & 150 & 31$\pm$16 \\
 5112730 &  4.56 &  320.0 & 0.25 &  45 & 56$\pm$18 \\
 5112938 &  4.73 &  320.0 & 0.30 &  65 & 45$\pm$11 \\
 5112950 &  4.35 &  319.5 & 0.38 &  38 & 61$\pm$18 \\
 5112974 &  4.32 &  309.6 & 0.24 &  60 & 50$\pm$12 \\
 5113441$^c$ & 11.75 &   89.0 & 0.13 & 730 & 18$\pm$18 \\
 5200152 &  4.73 &  327.2 & 0.28 &  50 & 70$\pm$15 \\
  \hline
\end{tabular}

\scriptsize{
$^a$: KIC 4937257 and KIC 4937775 are absent in \cite{2017NatAs...1E..64C}.

$^b$: $\times$ symbols indicate the absence of any reliable asymptotic fit for KIC 4937770.

$^c$: Different solutions in $\Tg$ are possible for KIC 5112361, which all provide a high inclination. Different solutions in $\dnurot$ are possible for KIC 5113441, which all provide a low inclination.

Typical uncertainties for those stars with low S/R spectra are 0.7\,\% in $\Dnu$ and $\Tg$, 12\,\% in $q$,  and 8\,\% in $\dnurot$.

}
\end{table}

We measured inclinations of red giants in the open clusters  NGC 6819 observed by \Kepler\ \citep[e.g.,][]{2011ApJ...729L..10B,2011ApJ...739...13S,2012MNRAS.419.2077M}. We selected the stars that exhibit mixed modes and could fit 20 mixed-mode patterns with the asymptotic expansion. In one case, the asymptotic fit is impossible, due to a low S/R. In two other cases, different possible solutions exist, based either on different period spacings, or on different rotational splittings, but without any ambiguity for the inclination measurement: when two peaks dominate per period spacing, the inclination is necessarily high, whereas it is low when one single peak only is present. We completed this list with other NGC 6819 members listed in \cite{2017MNRAS.472..979H} and could fit two additional stars, which incidentally show a large inclination. Results for the inclinations and rotational splittings are given in Table \ref{tab-NGC6819-incli}. As shown in Fig.~\ref{fig-i-histo}, the distribution of the stellar inclinations mimics the $\sin i$ relation expected for random inclinations, except near 90$^\circ$, due to bias mentioned above. A similar test performed on the open cluster NGC 6791 reaches the same conclusion.

\begin{figure}
\includegraphics[width=8.8cm]{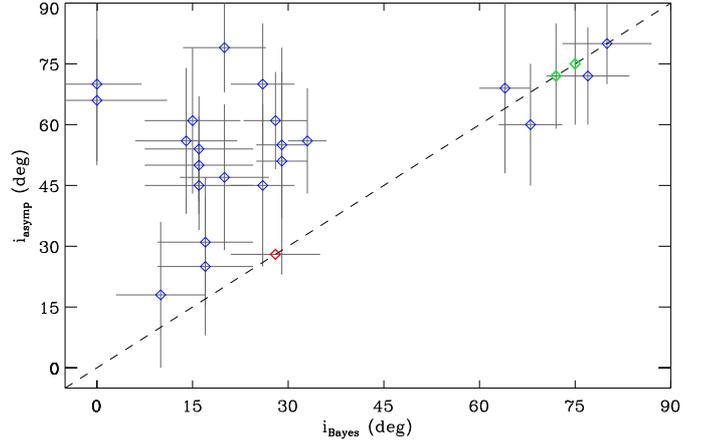}
\caption{Comparison of the inclinations of the spin axis of the
stars in NGC 6819. Inclinations measured by \cite{2017NatAs...1E..64C} are plotted on the x-axis, while
inclinations derived from the asymptotic analysis are on the y-axis; 1-$\sigma$ uncertainties are indicated by vertical and horizontal error bars. The red symbol shows the case where no asymptotic fit could be found, and the green ones to cases without Bayesian fit. The dashed line corresponds to the 1:1 relation.}
\label{fig-i-amas}
\end{figure}

Low stellar inclinations in NGC 6819 and 6791 were measured by \cite{2017NatAs...1E..64C}, using a Bayesian analysis, from which aligned spins were inferred. Our measurements however contradict their claim, as shown in Fig.~\ref{fig-i-amas} for NGC 6819. In fact, our measurements compared to theirs agree for high inclinations, whereas they mostly disagree for low inclinations. Comparison with the asymptotic fits shows that their Bayesian rotational splittings are most often overestimated and that the related inclinations are most often underestimated.

\begin{figure}[t]
\includegraphics[width=8.8cm]{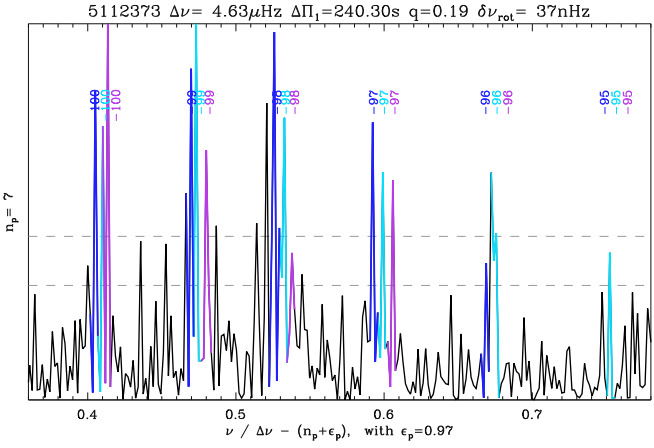}
\caption{Fit of the mixed modes corresponding to $\np=7$ in KIC 5112373 (NGC 6819 member). The color codes the azimuthal order: $m=+1$ in purple, $m=-1$ in blue. The gray dashed lines indicate the two thresholds used in this work, corresponding to height-to-background ratios of 7 and 10. Contrary to the analysis conducted by \cite{2017NatAs...1E..64C}, modes with $m=\pm1$ are clearly identified.}
\label{fig-zoom-5112373}
\end{figure}

In fact, the asymptotic fit can be used as a prior for the Bayesian fit. It indicates that the rotational splitting is derived from the thin gravity-dominated mixed modes, with narrow widths   (Eq.~\ref{eqt-largeur}) and average rotational splittings slightly inferior to the mean core rotation $\dnurot$ (Eq.~\ref{eqt-zeta-rotation2}). Mixed modes at low pressure radial orders, with frequencies much below $\numax$, are especially informative, since previous work has shown that their radial mode widths, hence their mixed-mode widths according to Eq.~(\ref{eqt-largeur}), are the thinnest possible  \citep[Fig. 5 of][]{2018arXiv180503690V}. Figure 1 of the supplementary
material of \cite{2017NatAs...1E..64C} provides an explanation of the discrepant Bayesian values. Their fit of the star KIC 5112373 in NGC 6819 provides nearly uniform large mode widths, relevant for the pressure-dominated mixed modes but much too high for gravity-dominated modes, in contradiction with the physical variation indicated by  Eq.~(\ref{eqt-largeur}). As a consequence, their fit assumes that all the power is concentrated in the $m=0$ mode; the resulting stellar inclination is $20\pm8^\circ$.
We show the asymptotic solution of KIC 5112373 in Fig.~\ref{fig-zoom-5112373}, with thin gravity dominated mixed modes and the clear identification of triplets. Since $m=\pm1$ modes are observed all along the spectrum, our solution for the inclination is larger, about $47\pm18^\circ$.

\begin{figure}[t]
\includegraphics[width=8.8cm]{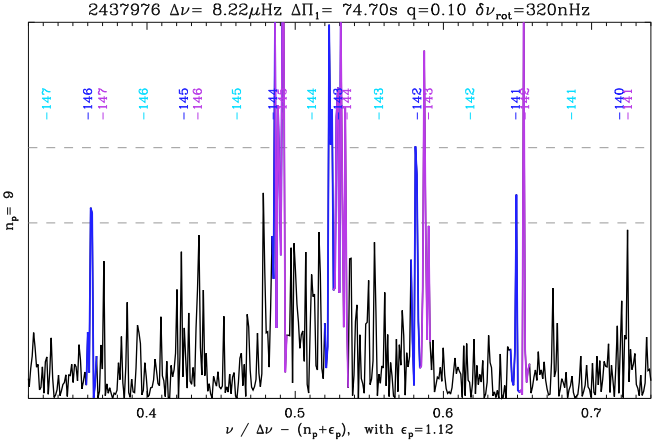}
\caption{Fit of the mixed modes corresponding to $\np=9$ in KIC
2437976 (NGC 6791 member). The color codes the azimuthal order: $m=+1$ in purple, $m=-1$ in blue. The location of $m=0$ modes is indicated in light blue, but none shows a large height for this star seen equator-on. The gray dashed lines indicate the two thresholds used in this work, corresponding to height-to-background ratios of 7 and 10. Many peaks above the threshold value 5.5 that rejects the null hypothesis at the 5\,{\%}-level follow the mixed-mode pattern.}
\label{fig-zoom-2437976}
\end{figure}

We provide another example with the star KIC 2437976, a NGC 6791 member. As shown in Fig.~\ref{fig-zoom-2437976}, rotational splittings are explained in a consistent way with thin unresolved gravity-dominated mixed modes and a rotation rate rapid enough to ensure that close modes \emph{do not} belong to the same multiplets. All peaks can be explained by the $m=\pm 1$ modes. In practice, $m=0$ modes
are absent, so that this star has necessarily an inclination close to $90^\circ$, whereas \cite{2017NatAs...1E..64C} measured $i \simeq 0^\circ$. We conclude that some of the low inclinations reported in \cite{2017NatAs...1E..64C} are incompatible with the analysis presented here. It seems that the difference is due to a too low range of the linewidth priors in the Bayesian analysis, which favors a solution with a low inclination angle and a high splitting. As a result, stellar spins in old open clusters are neither aligned nor quasi parallel to the line of sight. Our study emphasizes a major role for the asymptotic analysis: providing relevant estimates of all features of the mixed-mode pattern, including mode widths.

\section{Observability of the mixed modes\label{application-observability}}

All the information derived from mixed modes relies on their observability. The properties of the function $\zeta$ can be used to assess under which conditions mixed modes can be actually observed. To achieve this, we investigate first the domain where pressure-dominated mixed modes are observed, then the condition for observing gravity-dominated mixed modes.

\subsection{Pressure-dominated mixed modes}

We can define the frequency range where mixed modes are pressure-dominated (pm) from the full width at half minimum of the $\zeta$ function. So, these modes cover a range, expressed in terms of the pressure phase $\thetap$ (Eq.~\ref{eqt-zeta}), verifying
\begin{equation}\label{eqt-largeur-pm}
  \left.\delta\thetap\right|\ind{pm} = 2 q\; \sqrt{1+{1\over \dens q}}
\end{equation}
under the assumption that $q$ is small, which is verified for all stars except at the transition between subgiants and red giants \citep{2017A&A...600A...1M}. When expressed in frequency and compared to the large separation, this condition corresponds to a frequency range surrounding each pure pressure modes with a width $\delta\nu\ind{pm}$ defined by
\begin{equation}\label{eqt-largeur-pm2}
  {\delta\nu\ind{pm} \over \Dnu} =  {2 q\ \over \pi}\; \sqrt{1+{1\over \dens q}}
  .
\end{equation}
The variations in $q$ and $\dens$ explain the narrowing of the region with
pressure-dominated mixed modes when stars evolve on the RGB. An example is shown in the Appendix (Fig.~\ref{fig-fit-2443903}). The expression of $\delta\nu\ind{pm}$
also shows that red-clump stars, with larger $q$ show
pressure-dominated mixed modes in a broader region than RGB stars.

\subsection{Visible gravity-dominated mixed modes}

The non-dilution of the mode height expressed by Eq.~\ref{eqt-zeta-H} can  be used to define a criterion of visibility of the gravity-dominated (gm) mixed modes. So, they are clearly
visible when they show  heights similar to those of the pressure modes ($H_\nm=H_0$), hence when $\Gamma_0 (1-\zeta) \ge 2 \dfres / \pi$ (Eq.~\ref{eqt-largeur}). This condition translates into
\begin{equation}\label{eqt-condi-visi-gm}
  (1-q^2)\; \sin^2\thetap \le {q \over \dens} \left({\pi\over 2} {\Gamma_0 \over \dfres} -1\right) - q^2
  .
\end{equation}
Except at the transition from subgiants to red giants, where mixed
modes are unambiguously visible \citep{2013ApJ...767..158B,
2014A&A...564A..27D}, the terms $q^2$ are negligible, so that
modes are clearly visible if
\begin{equation}\label{eqt-condi-visi-gm2}
   |\sin \thetap |\ind{gm}
   \le
   \sqrt{{q \over \dens}
   \left( {\pi\over 2} {\Gamma_0 \over \dfres} -1\right)}
  .
\end{equation}
This condition for observing gravity-dominated mixed modes has many consequences:\\
- It can be fulfilled only if the definition of the right term is ensured, which requires a frequency resolution low enough compared to the radial mode width. With $\Gamma_0$ in the range [100,\,150\,nHz], the observation must last 50-75~days at least. In fact, mixed modes were observable with CoRoT runs lasting about
150 days \citep{2011A&A...532A..86M}, but are hardly observable with K2 80-day time series \citep{2017ApJ...835...83S}.\\
- When stars evolve on the RGB, the decrease in $q$ and increase in $\dens$ contribute to the narrowing of observable modes. Mixed modes are more easily visible in the red clump, owing to larger $q$ values. This criterion is implicitly used by \cite{2017MNRAS.466.3344E} for their determination of the evolutionary state of red-giant stars.
\\
- All mixed modes are clearly visible when the condition expressed by Eq.~(\ref{eqt-condi-visi-gm2}) is always met, that is when $\dens \le q (\pi\Gamma_0/2\dfres -1)$. This condition is met for subgiants, on the lower RGB, and for secondary-clump stars
\citep{2014A&A...572L...5M}.\\
- No mixed mode can be observed when the condition is so drastic that only pressure-dominated mixed modes can be observed. The combination of the conditions expressed by Eq.~(\ref{eqt-largeur-pm}) and Eq.~(\ref{eqt-condi-visi-gm2}) yields the limit of visibility of gravity-dominated mixed modes, expressed by a condition on the mixed-mode density
\begin{equation}\label{eqt-limite-visi-gm}
    \dens
    \le
    {1\over 4 q} \left({\pi\over 2} {\Gamma_0 \over \dfres} - 5 \right)
    .
\end{equation}
In the conditions of observation of \Kepler, with typical parameters defined as in \cite{2017A&A...598A..62M}, this limit corresponds to a mode density $\dens$ of about 25, for RGB and clump stars, over which no gravity-dominated mixed modes can be identified. This theoretical estimate is observed in practice, with a few exceptions with larger $\dens$ (Fig.~\ref{fig-HR}). On the RGB, observation of mixed modes with \Kepler\ is limited to $\Dnu\ge 6\,\mu$Hz, whereas the limit is around 3\,$\mu$Hz for clump stars. As a consequence, visible mixed modes in an oscillation spectrum with $\Dnu$ in the range [3,\,6\,$\mu$Hz] most often indicate a red-clump star. Incidentally, the location of the RGB bump was recently identified by \cite{2018ApJ...859..156K} in the range [5,\,6\,$\mu$Hz], depending on the stellar mass and metallicity. This means that sounding the bump with mixed modes will be very difficult, if not impossible.
\section{Conclusion\label{conclusion}}

The asymptotic analysis allows us to depict the whole properties of the mixed-mode spectrum in a consistent way. Period spacings, rotational splittings, mode widths, and mode heights, all depend on the mode inertia so that all are related to the parameter $\zeta$. We could derive interesting properties:

- The asymptotic fit of the mixed modes proves to be precise and unbiased. Its precision for the RGB stars is so high that the asymptotic expansion of gravity modes can be validated when buoyancy glitches are absent. This ensures the delivery of accurate asymptotic parameters $\Tg$, $q$, and $\epsg$. We found only one RGB star with such buoyancy glitches; on the contrary, buoyancy glitches are often present in red-clump stars.

- The period spacings and rotational splittings are better estimated with integrated values of the function
$\zeta$. The use of these mean values $\zmoyn$ and $\zmoym$ is useful for evolved RGB stars and is mandatory for stars with intricate splittings and spacings. Using the stretched period \citep{2015A&A...584A..50M} is in fact equivalent.

- The gravity asymptotic parameters $\Tg$ and $\epsg$ can now be accurately determined, with typical accuracy of respectively 0.06\,s  and 0.1 on the RGB, and 0.22\,s and 0.08 in the red clump. This opens the way to a fruitful dialogue with theoretical developments \citep{2006PASJ...58..893T,2016PASJ...68...91T,2016PASJ...68..109T} and modeling \citep[e.g.,][]{2015MNRAS.453.2290B,2015ApJ...805..127C}.

- We have made clear that observing mixed modes in evolved red giants requires an observation duration longer than $\simeq 100\,$days. However, gravity-dominated mixed modes are no longer observable when the stars are more evolved than $\Dnu \simeq 6\,\mu$Hz on the RGB, or $\Dnu \simeq 3\,\mu$Hz in the red clump. These thresholds are indicative values: the natural spread of the seismic parameters with respect to their mean values explain slight differences.

- We have demonstrated the non-alignment of the rotation axis of the stars belonging to the old open clusters NGC 6791 and NGC 6819. These results contradict previous findings by  \cite{2017NatAs...1E..64C} and illustrate how useful the asymptotic fit will be in the future when used to define priors to any Bayesian or other type of fit of mixed modes.
\begin{acknowledgements}
We thank the entire \Kepler\ team, whose efforts made these results possible.
BM warmly thanks Yvonne Elsworth, James Kuszlewicz and Masao Takata for their comments on the draft submitted for internal review on the website of the \Kepler\ Asteroseismic Science Operations Center.
We acknowledge financial support from the Programme National de
Physique Stellaire (CNRS/INSU). BM acknowledges the support of the International Space Institute (ISSI) for the program  AsteroSTEP (Asteroseismology of STEllar Populations)
\end{acknowledgements}

\bibliographystyle{aa} 

\begin{appendix}

\section{Seismic parameters\label{appendix-6144777}}

We used  KIC 6144777 as a case study (Fig.~\ref{fig-fit-6144777}).
Table~\ref{table-freq6144777} provides the fit of its radial dipole mixed modes. Our results are in agreement with those published by \cite{2015A&A...579A..83C} and derive a similar number of modes (about 100), but also show differences:\\
- The determination of the frequencies in \cite{2015A&A...579A..83C} can be as precise as 0.3\,nHz. This precision of about $\dfres/30$ was corrected into about $\dfres/10$ in their corrigendum \citep{2018A&A...612C...2C}, which remains surprisingly good; the frequencies we obtain are given with a precision that is at best about half the frequency resolution ($\simeq$4\,nHz).\\
- Their mode widths are quite different and, most often, larger than ours;\\
- Heights also differ, which can come from a different treatment of the time series.\\
A large agreement is also met with the results obtained by \cite{2018MNRAS.476.1470G} with a peak detection algorithm that works in a fully blind manner, if we relax their uncertainties that can be as low as $\dfres/20$.

The potential of the comparison between methods based on different principles is very high: coupling the physics of the asymptotic expansion and the power of a pure numerical approach is the next step for delivering duly identified mixed modes.\\

\begin{figure*}
\includegraphics[width=15cm]{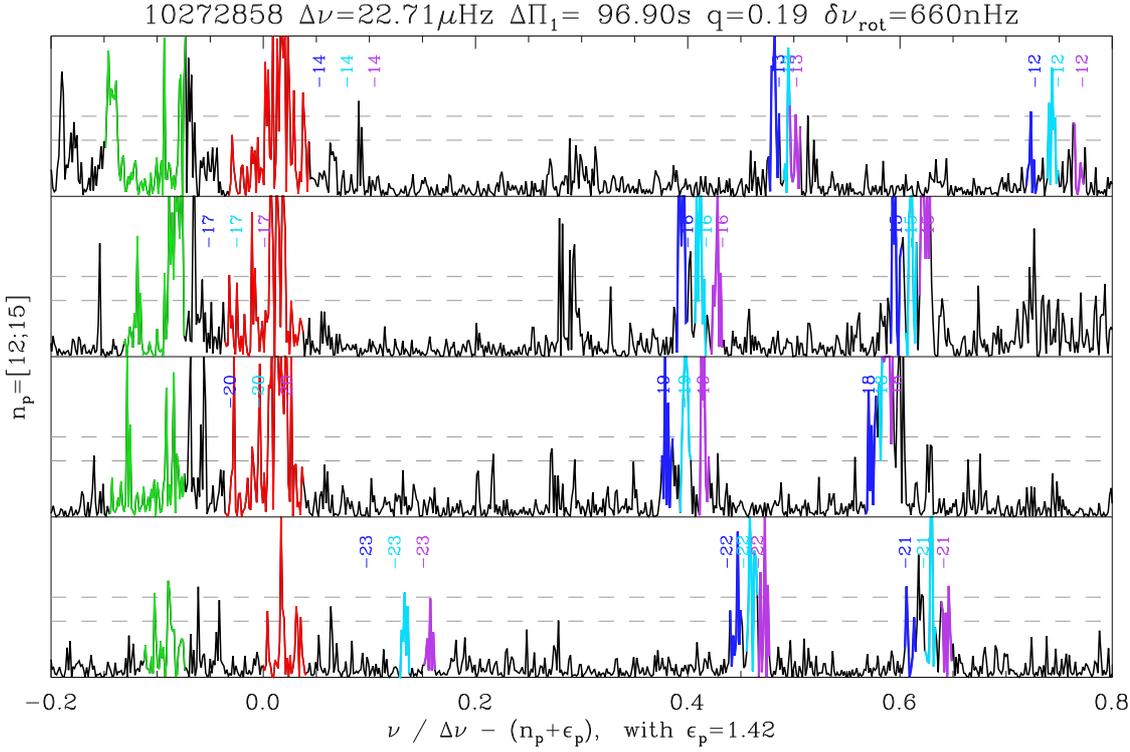}
\caption{Fit of the oscillation pattern of the low RGB star KIC 10272858, at the limit of validity of the asymptotic pattern. Owing to the small radial orders, small shifts are seen between observed and asymptotic spectra. Same style as Fig.~\ref{fig-fit-6144777}, but $\ell=3$ modes appear near the abscissa 0.28}\label{fig-fit-10272858}
\end{figure*}

\begin{figure*}
\includegraphics[width=15cm]{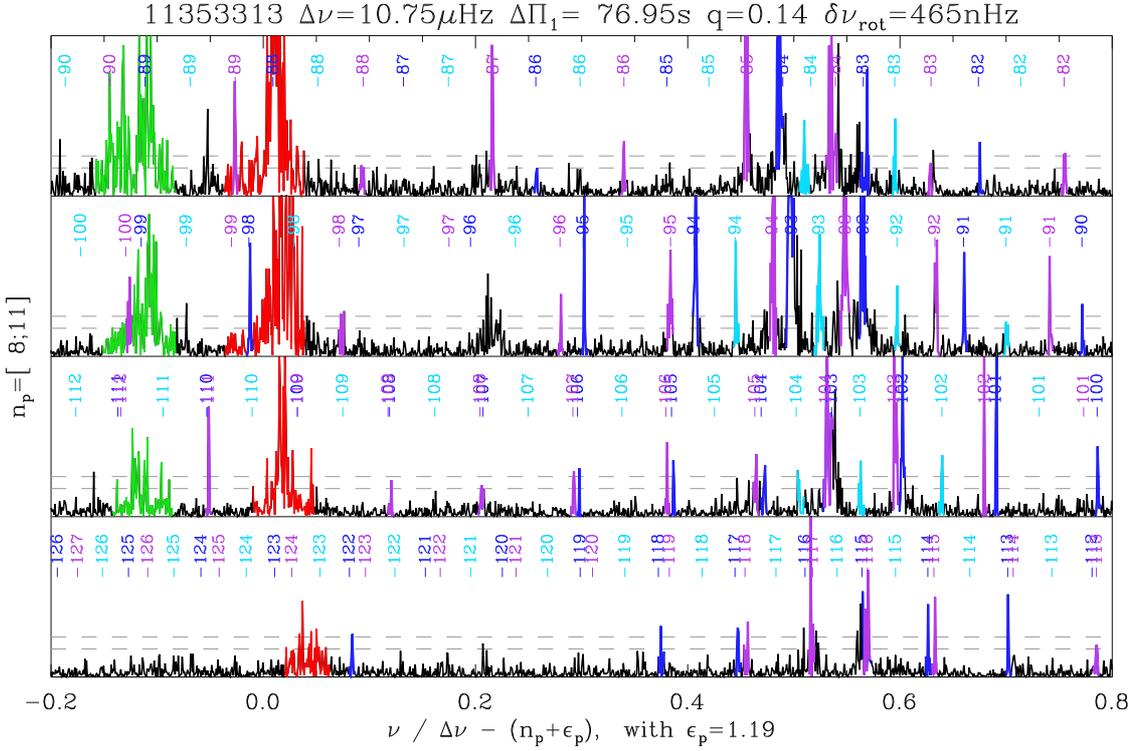}
\caption{Fit of the oscillation pattern of the RGB star KIC 11353313. Same style as Fig.~\ref{fig-fit-6144777}.}\label{fig-fit-11353313}
\end{figure*}

\begin{figure*}
\includegraphics[width=15cm]{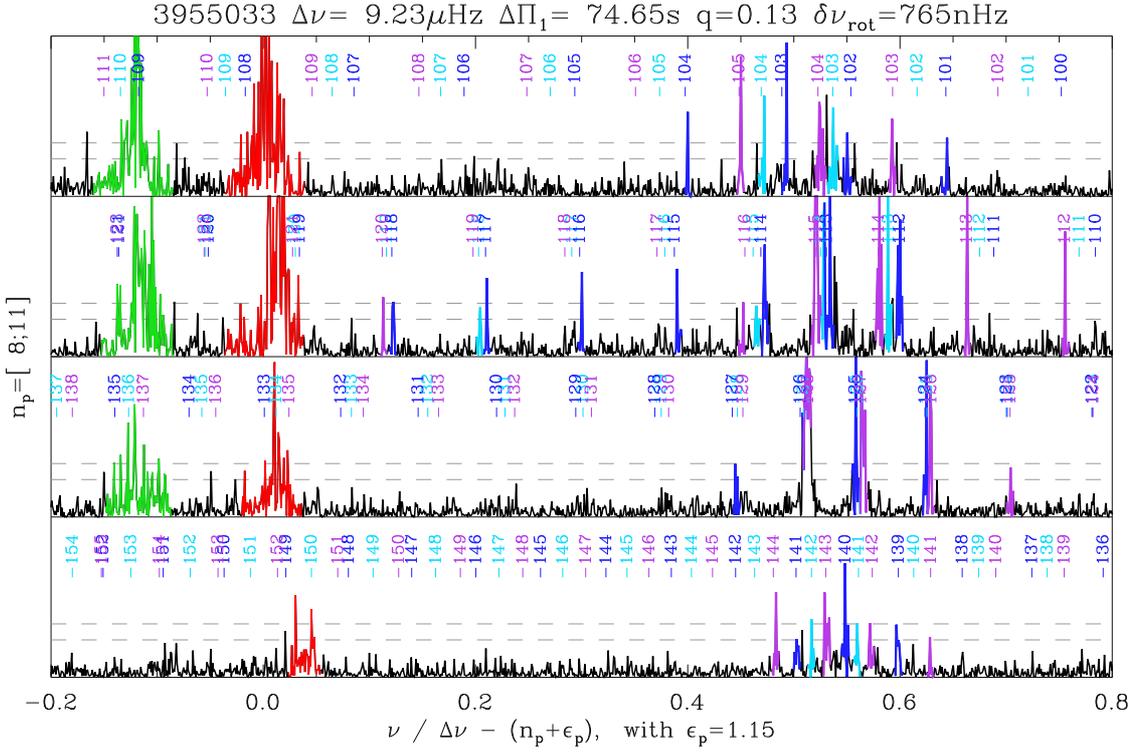}
\caption{Fit of the oscillation pattern of the RGB star KIC 3955033. The overlap of mixed modes with different mixed-mode orders is the signature of the rapid core rotation. The second rotation crossing, where all components of the multiplets overlap \citep{2017EPJWC.16004005G}, occurs at the mixed-order $\nm = -122$. Same style as Fig.~\ref{fig-fit-6144777}.}\label{fig-fit-3955033}
\end{figure*}

\begin{figure*}
\includegraphics[width=15cm]{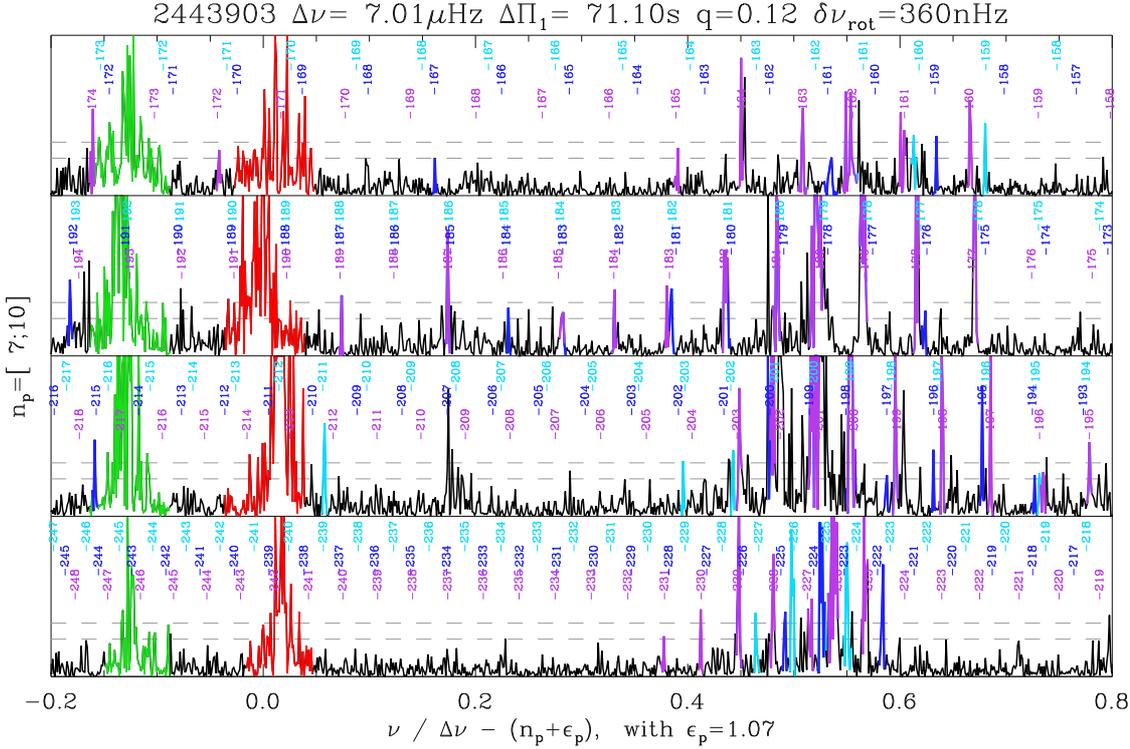}
\caption{Fit of the oscillation pattern of the evolved RGB star KIC
2443903, near the limit of capability of identification, with a large crowding due to the high mode density. The second rotation crossing, where all $m$ components apparently coincide, occurs at $\nm=-189$ (with an abscissa $\simeq 0.1$ and $\np=9$); the third crossing, where $|m|=1$ components apparently coincide with $m=0$ inbetween, occurs at $\nm=-233$ (with an abscissa $\simeq 0.25$ and $\np=7$). Same style as Fig.~\ref{fig-fit-6144777}. Note that the modes with large heights at an abscissa $\simeq 0.2$  are $\ell=3$ modes.}
\label{fig-fit-2443903}
\end{figure*}

\begin{figure*}
\includegraphics[width=15cm]{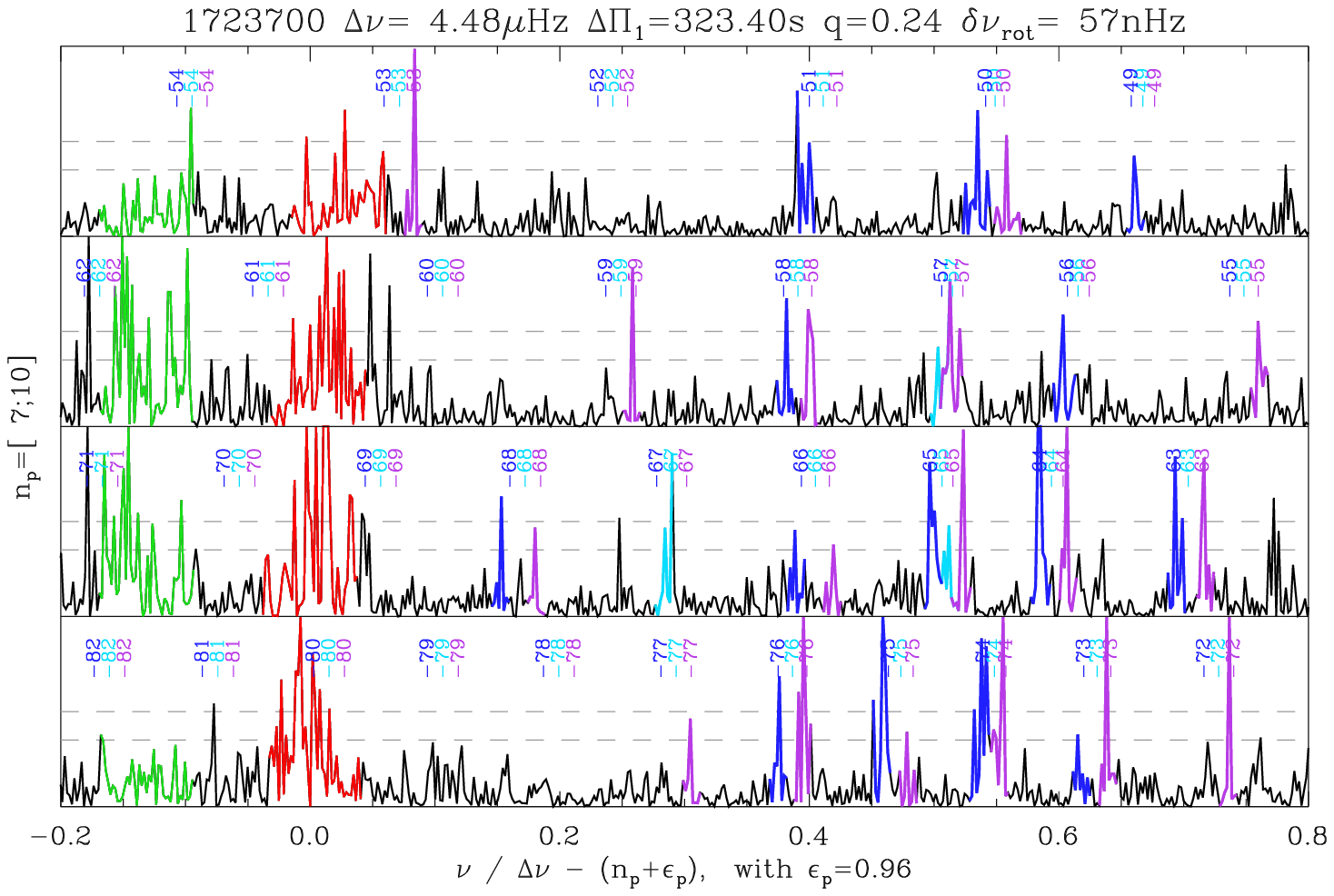}
\caption{Fit of the oscillation pattern of the red-clump star KIC 1723700. Buoyancy glitches explain the small shifts between observed and asymptotic spectra but do no hamper the mode identification. Same style as Fig.~\ref{fig-fit-6144777}.}\label{fig-fit-1723700}
\end{figure*}

\begin{figure*}
\includegraphics[width=15cm]{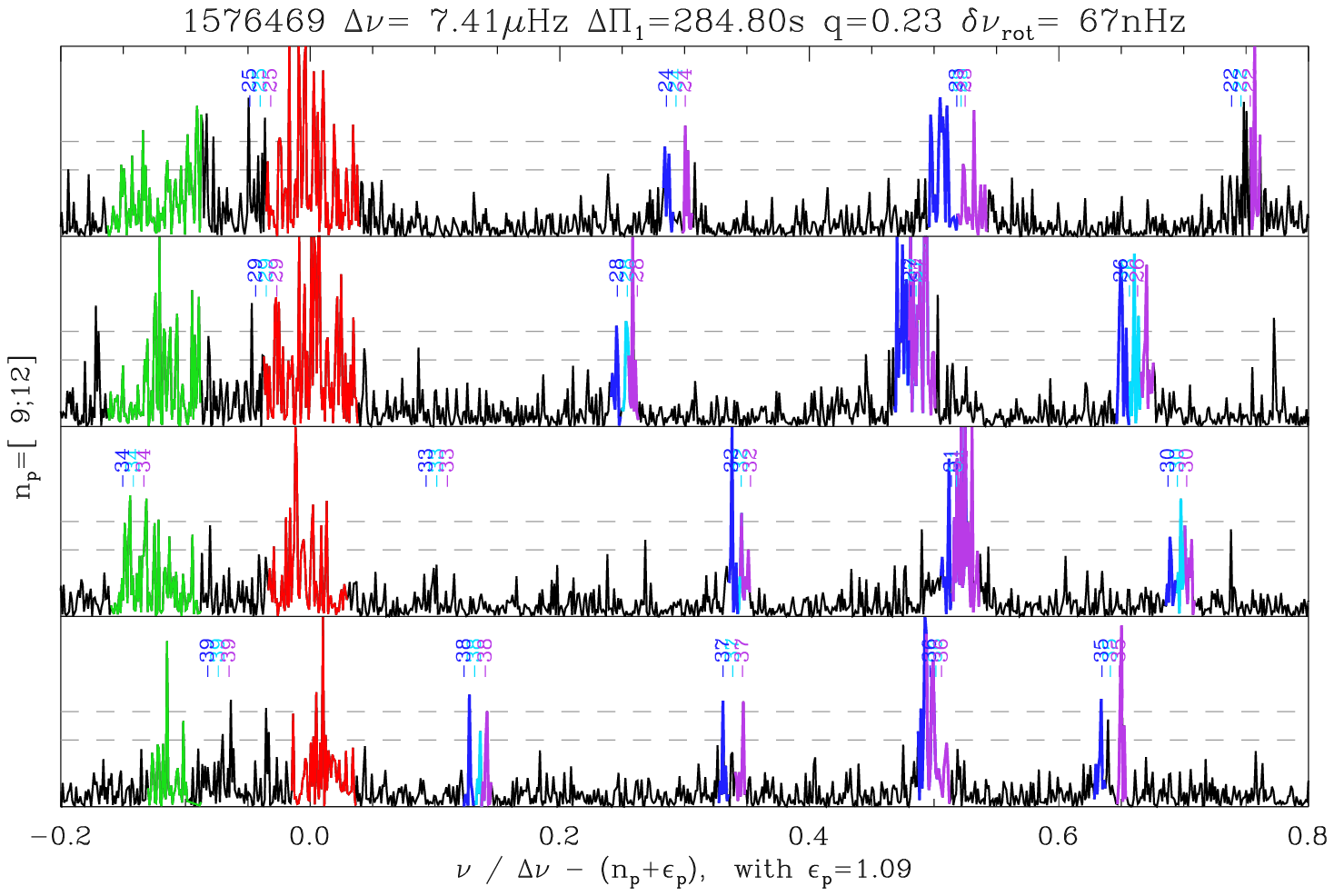}
\caption{Fit of the oscillation pattern of the secondary-clump star KIC 1725190. Buoyancy glitches explain the small shifts between observed and asymptotic spectra but do no hamper the mode identification. Same style as
Fig.~\ref{fig-fit-6144777}.}\label{fig-fit-1725190}
\end{figure*}

The \'echelle diagrams of the stars mentioned in the main text are also presented:\\
- KIC 10272858 lies on the low part of the RGB (Fig.~\ref{fig-fit-10272858});\\
- KIC 11353313 is on the RGB (Fig.~\ref{fig-fit-11353313});\\
- KIC 3955033 is a RGB star with a rapid core rotation (Fig.~\ref{fig-fit-3955033}); its frequencies are given in Table \ref{table-freq3955033}; \\
- KIC 2443903 is more evolved on the RGB, at the limit of detection of mixed modes (Fig.~\ref{fig-fit-2443903});\\
- KIC 1723700 is in the red clump star (Fig.~\ref{fig-fit-1723700});\\
- and KIC 1725190 is a secondary red clump star (Fig.~\ref{fig-fit-1725190}). \\

\longtab{1}{\tiny
\begin{longtable}{rrrcccccccc}
\caption{Oscillation pattern of the RGB star KIC 6144777\label{table-freq6144777}}\\
\hline
   $\np$ & $\nm$ & $m$ & $\zeta$&$\nu\ind{as}$ & $\nu$   &$x$ &$\Gamma\ind{as}$& $\Gamma$& $H$  & $R$\\
         &       &     &        &($\mu$Hz)     &($\mu$Hz)&    &  ($\mu$Hz) & ($\mu$Hz)   &(ppm$^2\mu$Hz$^{-1}$)         \\
\hline
\endfirsthead
\caption{continued.}\\
\hline
  $\np$ & $\nm$ & $m$ & $\zeta$&$\nu\ind{as}$ & $\nu$   & $x$ & $\Gamma\ind{as}$& $\Gamma$& $H$  & $R$\\
        &       &     &        &($\mu$Hz)     &($\mu$Hz)&     &($\mu$Hz) & ($\mu$Hz)   &(ppm$^2\mu$Hz$^{-1}$)         \\
\hline
\endhead
\hline
\endfoot
 \multicolumn{11}{c}{Radial modes}\\
  \hline
 8 &  &  &  & 101.916 &$101.916\pm 0.014$&$ 0.040$&  &$ 0.120\pm0.023$&$  1390\pm   316$&  24.1 \\
 9 &  &  &  & 112.612 &$112.612\pm 0.011$&$ 0.010$&  &$ 0.106\pm0.016$&$  3314\pm   497$&  85.2 \\
10 &  &  &  & 123.726 &$123.622\pm 0.006$&$ 0.008$&  &$ 0.054\pm0.007$&$ 23358\pm  3059$& 338.5 \\
11 &  &  &  & 134.574 &$134.534\pm 0.008$&$-0.003$&  &$ 0.087\pm0.012$&$ 10451\pm  1305$& 195.7 \\
12 &  &  &  & 145.842 &$145.585\pm 0.017$&$-0.002$&  &$ 0.244\pm0.057$&$  1491\pm   441$&  26.6 \\
13 &  &  &  & 156.846 &$156.846\pm 0.042$&$ 0.019$&  &$ 0.258\pm0.074$&$   317\pm   121$&  10.7 \\
  \hline
  \multicolumn{11}{c}{Dipole mixed modes}\\
  \hline
 8 &$-112$&$ 1$& 0.9522 & 105.827 &$105.836\pm 0.007$&$ 0.396$&  0.009 &$ 0.007\pm0.004$&$  1185\pm   281$&  10.1 \\
 8 &$-111$&$ 1$& 0.8506 & 106.633 &$106.648\pm 0.007$&$ 0.469$&  0.027 &$ 0.011\pm0.004$&$  2124\pm   440$&  26.0 \\
 8 &$-110$&$-1$& 0.7311 & 106.957 &$106.985\pm 0.014$&$ 0.500$&  0.048 &$ 0.068\pm0.017$&$   454\pm   149$&  14.5 \\
 8 &$-110$&$ 0$& 0.6467 & 107.134 &$107.137\pm 0.010$&$ 0.514$&  0.064 &$ 0.042\pm0.009$&$   658\pm   145$&  13.9 \\
 8 &$-110$&$ 1$& 0.5989 & 107.278 &$107.322\pm 0.008$&$ 0.530$&  0.072 &$ 0.040\pm0.007$&$  1497\pm   250$&  30.6 \\
 8 &$-109$&$-1$& 0.6428 & 107.551 &$107.604\pm 0.009$&$ 0.556$&  0.064 &$ 0.055\pm0.011$&$  1272\pm   292$&  17.8 \\
 8 &$-109$&$ 0$& 0.7161 & 107.706 &$107.730\pm 0.010$&$ 0.567$&  0.051 &$ 0.036\pm0.008$&$   630\pm   157$&  11.2 \\
 8 &$-109$&$ 1$& 0.7983 & 107.899 &$107.909\pm 0.007$&$ 0.584$&  0.036 &$ 0.018\pm0.005$&$  1646\pm   323$&  18.0 \\
 8 &$-108$&$-1$& 0.8911 & 108.264 &$108.263\pm 0.010$&$ 0.616$&  0.020 &$ 0.016\pm0.005$&$   770\pm   254$&  12.2 \\
 8 &$-108$&$ 0$& 0.9206 & 108.482 &$108.484\pm 0.006$&$ 0.636$&  0.014 &$ 0.007\pm0.004$&$  1554\pm   335$&  14.0 \\
 8 &$-108$&$ 1$& 0.9406 & 108.710 &$108.700\pm 0.011$&$ 0.655$&  0.011 &$ 0.014\pm0.005$&$   505\pm   186$&   8.8 \\
 8 &$-107$&$-1$& 0.9617 & 109.140 &$109.138\pm 0.007$&$ 0.695$&  0.007 &$ 0.006\pm0.004$&$  1303\pm   313$&  11.9 \\
 8 &$-107$&$ 0$& 0.9683 & 109.373 &$109.375\pm 0.008$&$ 0.716$&  0.006 &$ 0.005\pm0.004$&$   783\pm   323$&   7.2 \\
 8 &$-106$&$ 0$& 0.9822 & 110.304 &$110.274\pm 0.008$&$ 0.798$&  0.003 &$ 0.006\pm0.004$&$   803\pm   175$&   8.4 \\
 8 &$-104$&$ 0$& 0.9889 & 112.235 &$112.235\pm 0.007$&$-0.024$&  0.002 &$ 0.005\pm0.004$&$   959\pm   335$&   9.3 \\
 9 &$-103$&$-1$& 0.9889 & 112.989 &$113.018\pm 0.012$&$ 0.047$&  0.002 &$ 0.006\pm0.004$&$   739\pm   269$&   8.1 \\
 9 &$-103$&$ 0$& 0.9889 & 113.228 &$113.226\pm 0.006$&$ 0.065$&  0.002 &$ 0.005\pm0.004$&$  1736\pm   480$&  17.3 \\
 9 &$-102$&$-1$& 0.9872 & 113.999 &$113.996\pm 0.007$&$ 0.135$&  0.002 &$ 0.006\pm0.004$&$   789\pm   174$&   8.2 \\
 9 &$-102$&$ 1$& 0.9856 & 114.477 &$114.478\pm 0.007$&$ 0.179$&  0.003 &$ 0.005\pm0.004$&$   919\pm   297$&   9.4 \\
 9 &$-101$&$ 0$& 0.9800 & 115.262 &$115.254\pm 0.008$&$ 0.249$&  0.004 &$ 0.005\pm0.004$&$   701\pm   221$&   7.3 \\
 9 &$-100$&$-1$& 0.9672 & 116.058 &$116.058\pm 0.006$&$ 0.322$&  0.006 &$ 0.006\pm0.004$&$  1721\pm   254$&  19.0 \\
 9 &$-100$&$ 0$& 0.9600 & 116.292 &$116.285\pm 0.007$&$ 0.343$&  0.007 &$ 0.006\pm0.004$&$  1150\pm   273$&  12.1 \\
 9 &$-100$&$ 1$& 0.9506 & 116.523 &$116.525\pm 0.006$&$ 0.364$&  0.009 &$ 0.006\pm0.004$&$  2571\pm   532$&  27.2 \\
 9 &$ -99$&$-1$& 0.9056 & 117.072 &$117.076\pm 0.009$&$ 0.414$&  0.017 &$ 0.022\pm0.006$&$   459\pm   134$&  12.7 \\
 9 &$ -99$&$ 0$& 0.8694 & 117.291 &$117.318\pm 0.025$&$ 0.436$&  0.023 &$ 0.040\pm0.013$&$    77\pm    34$&   7.2 \\
 9 &$ -99$&$ 1$& 0.8189 & 117.489 &$117.520\pm 0.006$&$ 0.455$&  0.033 &$ 0.011\pm0.004$&$  4320\pm   641$&  54.2 \\
 9 &$ -98$&$-1$& 0.6256 & 117.926 &$117.973\pm 0.007$&$ 0.496$&  0.067 &$ 0.043\pm0.007$&$  2105\pm   327$&  32.6 \\
 9 &$ -98$&$ 0$& 0.5622 & 118.077 &$118.138\pm 0.009$&$ 0.511$&  0.079 &$ 0.033\pm0.007$&$  5534\pm  1222$&  91.7 \\
 9 &$ -98$&$ 1$& 0.5461 & 118.211 &$118.282\pm 0.009$&$ 0.524$&  0.082 &$ 0.034\pm0.007$&$  3206\pm   652$&  69.8 \\
 9 &$ -97$&$-1$& 0.6989 & 118.597 &$118.638\pm 0.008$&$ 0.556$&  0.054 &$ 0.034\pm0.007$&$  3294\pm   723$& 142.5 \\
 9 &$ -97$&$ 0$& 0.7744 & 118.766 &$118.815\pm 0.007$&$ 0.572$&  0.041 &$ 0.013\pm0.004$&$  3831\pm   566$&  66.2 \\
 9 &$ -97$&$ 1$& 0.8394 & 118.969 &$118.994\pm 0.006$&$ 0.588$&  0.029 &$ 0.010\pm0.004$&$ 10109\pm  1478$& 154.7 \\
 9 &$ -96$&$-1$& 0.9267 & 119.532 &$119.539\pm 0.008$&$ 0.638$&  0.013 &$ 0.017\pm0.005$&$  1339\pm   307$&  23.5 \\
 9 &$ -96$&$ 0$& 0.9428 & 119.758 &$119.758\pm 0.006$&$ 0.658$&  0.010 &$ 0.008\pm0.004$&$  4038\pm   588$&  46.3 \\
 9 &$ -96$&$ 1$& 0.9539 & 119.989 &$119.991\pm 0.007$&$ 0.679$&  0.008 &$ 0.008\pm0.004$&$  2271\pm   444$&  25.7 \\
 9 &$ -95$&$-1$& 0.9717 & 120.623 &$120.617\pm 0.006$&$ 0.735$&  0.005 &$ 0.006\pm0.004$&$  3853\pm   586$&  44.1 \\
 9 &$ -95$&$ 1$& 0.9778 & 121.096 &$121.088\pm 0.006$&$ 0.778$&  0.004 &$ 0.006\pm0.004$&$  6667\pm   971$&  77.7 \\
 9 &$ -94$&$-1$& 0.9828 & 121.763 &$121.749\pm 0.006$&$-0.162$&  0.003 &$ 0.006\pm0.004$&$  5234\pm   769$&  61.2 \\
 9 &$ -93$&$-1$& 0.9867 & 122.933 &$122.925\pm 0.010$&$-0.055$&  0.002 &$ 0.007\pm0.004$&$  2021\pm   647$&  27.2 \\
10 &$ -92$&$ 0$& 0.9861 & 124.367 &$124.355\pm 0.006$&$ 0.074$&  0.002 &$ 0.006\pm0.004$&$  1007\pm   169$&  13.4 \\
10 &$ -91$&$-1$& 0.9833 & 125.344 &$125.329\pm 0.007$&$ 0.162$&  0.003 &$ 0.005\pm0.004$&$  2714\pm   740$&  33.8 \\
10 &$ -91$&$ 0$& 0.9817 & 125.582 &$125.568\pm 0.006$&$ 0.184$&  0.003 &$ 0.005\pm0.004$&$  1574\pm   381$&  19.7 \\
10 &$ -91$&$ 1$& 0.9800 & 125.819 &$125.796\pm 0.006$&$ 0.205$&  0.004 &$ 0.006\pm0.004$&$  5347\pm   831$&  67.2 \\
10 &$ -90$&$-1$& 0.9706 & 126.575 &$126.561\pm 0.007$&$ 0.274$&  0.005 &$ 0.006\pm0.004$&$  1987\pm   407$&  25.3 \\
10 &$ -90$&$ 0$& 0.9661 & 126.810 &$126.799\pm 0.008$&$ 0.296$&  0.006 &$ 0.009\pm0.004$&$  2797\pm   714$&  35.8 \\
10 &$ -90$&$ 1$& 0.9594 & 127.042 &$127.032\pm 0.006$&$ 0.317$&  0.007 &$ 0.006\pm0.004$&$  5283\pm  1186$&  67.9 \\
10 &$ -89$&$-1$& 0.9167 & 127.792 &$127.789\pm 0.006$&$ 0.385$&  0.015 &$ 0.010\pm0.004$&$  2454\pm   424$&  31.9 \\
10 &$ -89$&$ 0$& 0.8900 & 128.014 &$128.018\pm 0.007$&$ 0.406$&  0.020 &$ 0.007\pm0.004$&$  1441\pm   297$&  18.8 \\
10 &$ -89$&$ 1$& 0.8528 & 128.220 &$128.231\pm 0.007$&$ 0.425$&  0.026 &$ 0.009\pm0.004$&$  6599\pm  1274$&  92.5 \\
10 &$ -88$&$-1$& 0.6144 & 128.822 &$128.875\pm 0.008$&$ 0.484$&  0.069 &$ 0.021\pm0.005$&$  6832\pm  1038$& 132.3 \\
10 &$ -88$&$ 0$& 0.5406 & 128.970 &$129.032\pm 0.008$&$ 0.498$&  0.083 &$ 0.043\pm0.007$&$  3925\pm   617$&  91.8 \\
10 &$ -88$&$ 1$& 0.5044 & 129.095 &$129.168\pm 0.009$&$ 0.510$&  0.089 &$ 0.034\pm0.007$&$  5624\pm  1181$& 130.5 \\
10 &$ -87$&$-1$& 0.6672 & 129.569 &$129.608\pm 0.009$&$ 0.550$&  0.060 &$ 0.049\pm0.010$&$  3066\pm   755$& 157.3 \\
10 &$ -87$&$ 0$& 0.7439 & 129.731 &$129.772\pm 0.007$&$ 0.565$&  0.046 &$ 0.014\pm0.004$&$ 12352\pm  1855$& 257.9 \\
10 &$ -87$&$ 1$& 0.8133 & 129.927 &$129.955\pm 0.006$&$ 0.582$&  0.034 &$ 0.011\pm0.004$&$ 16357\pm  2408$& 226.5 \\
10 &$ -86$&$-1$& 0.9294 & 130.677 &$130.670\pm 0.006$&$ 0.647$&  0.013 &$ 0.007\pm0.004$&$  4084\pm   855$&  55.9 \\
10 &$ -86$&$ 0$& 0.9433 & 130.903 &$130.896\pm 0.006$&$ 0.667$&  0.010 &$ 0.007\pm0.004$&$  7019\pm  1024$&  96.4 \\
10 &$ -86$&$ 1$& 0.9533 & 131.134 &$131.122\pm 0.006$&$ 0.688$&  0.008 &$ 0.006\pm0.004$&$ 10575\pm  2196$& 145.8 \\
10 &$ -85$&$-1$& 0.9728 & 131.986 &$131.962\pm 0.009$&$ 0.764$&  0.005 &$ 0.009\pm0.004$&$  1419\pm   431$&  19.8 \\
10 &$ -85$&$ 0$& 0.9756 & 132.221 &$132.196\pm 0.009$&$ 0.785$&  0.004 &$ 0.006\pm0.004$&$   530\pm   151$&   7.4 \\
10 &$ -84$&$ 0$& 0.9833 & 133.590 &$133.573\pm 0.006$&$-0.090$&  0.003 &$ 0.006\pm0.004$&$  3228\pm   566$&  46.4 \\
11 &$ -83$&$ 0$& 0.9844 & 134.993 &$134.973\pm 0.006$&$ 0.037$&  0.003 &$ 0.006\pm0.004$&$  4520\pm   719$&  66.5 \\
11 &$ -82$&$-1$& 0.9806 & 136.185 &$136.161\pm 0.006$&$ 0.144$&  0.004 &$ 0.006\pm0.004$&$  1267\pm   184$&  20.0 \\
11 &$ -81$&$-1$& 0.9639 & 137.632 &$137.609\pm 0.006$&$ 0.275$&  0.007 &$ 0.006\pm0.004$&$  1148\pm   170$&  18.7 \\
11 &$ -81$&$ 0$& 0.9578 & 137.866 &$137.837\pm 0.007$&$ 0.296$&  0.008 &$ 0.006\pm0.004$&$   900\pm   151$&  13.9 \\
11 &$ -81$&$ 1$& 0.9494 & 138.096 &$138.075\pm 0.006$&$ 0.318$&  0.009 &$ 0.011\pm0.004$&$  2426\pm   366$&  39.1 \\
11 &$ -80$&$-1$& 0.8644 & 139.037 &$139.015\pm 0.007$&$ 0.403$&  0.024 &$ 0.012\pm0.004$&$  2323\pm   388$&  36.5 \\
11 &$ -80$&$ 0$& 0.8183 & 139.247 &$139.230\pm 0.006$&$ 0.422$&  0.033 &$ 0.020\pm0.005$&$  2879\pm   427$&  50.3 \\
11 &$ -80$&$ 1$& 0.7594 & 139.431 &$139.438\pm 0.007$&$ 0.441$&  0.043 &$ 0.029\pm0.006$&$   942\pm   147$&  18.2 \\
11 &$ -79$&$-1$& 0.4672 & 140.063 &$140.116\pm 0.009$&$ 0.503$&  0.096 &$ 0.078\pm0.015$&$  2479\pm   552$& 102.8 \\
11 &$ -79$&$ 0$& 0.4628 & 140.175 &$140.185\pm 0.012$&$ 0.509$&  0.097 &$ 0.128\pm0.031$&$  1534\pm   491$&  74.4 \\
11 &$ -79$&$ 1$& 0.4939 & 140.295 &$140.352\pm 0.009$&$ 0.524$&  0.091 &$ 0.071\pm0.012$&$  1831\pm   330$&  49.4 \\
11 &$ -78$&$-1$& 0.8100 & 140.998 &$141.019\pm 0.006$&$ 0.585$&  0.034 &$ 0.013\pm0.004$&$  3675\pm   536$&  61.6 \\
11 &$ -78$&$ 0$& 0.8550 & 141.194 &$141.202\pm 0.006$&$ 0.601$&  0.026 &$ 0.009\pm0.004$&$  3583\pm   541$&  58.3 \\
11 &$ -78$&$ 1$& 0.8894 & 141.409 &$141.419\pm 0.009$&$ 0.621$&  0.020 &$ 0.027\pm0.007$&$  1200\pm   332$&  44.1 \\
11 &$ -77$&$-1$& 0.9561 & 142.436 &$142.432\pm 0.006$&$ 0.713$&  0.008 &$ 0.007\pm0.004$&$  1831\pm   358$&  30.4 \\
11 &$ -77$&$ 0$& 0.9617 & 142.668 &$142.663\pm 0.006$&$ 0.734$&  0.007 &$ 0.006\pm0.004$&$  1912\pm   278$&  33.3 \\
11 &$ -77$&$ 1$& 0.9667 & 142.902 &$142.895\pm 0.009$&$ 0.755$&  0.006 &$ 0.009\pm0.004$&$   434\pm   133$&   7.3 \\
12 &$ -74$&$-1$& 0.9778 & 147.312 &$147.297\pm 0.007$&$ 0.154$&  0.004 &$ 0.005\pm0.004$&$   528\pm   146$&   9.4 \\
12 &$ -74$&$ 0$& 0.9761 & 147.549 &$147.551\pm 0.007$&$ 0.177$&  0.004 &$ 0.006\pm0.004$&$   557\pm   102$&  11.1 \\
12 &$ -74$&$ 1$& 0.9739 & 147.785 &$147.778\pm 0.006$&$ 0.197$&  0.005 &$ 0.006\pm0.004$&$   665\pm   120$&  12.0 \\
12 &$ -73$&$-1$& 0.9511 & 148.996 &$148.980\pm 0.009$&$ 0.306$&  0.009 &$ 0.010\pm0.004$&$   747\pm   193$&  18.1 \\
12 &$ -73$&$ 0$& 0.9422 & 149.226 &$149.215\pm 0.007$&$ 0.327$&  0.010 &$ 0.009\pm0.004$&$   756\pm   160$&  13.9 \\
12 &$ -73$&$ 1$& 0.9306 & 149.452 &$149.441\pm 0.006$&$ 0.348$&  0.013 &$ 0.011\pm0.004$&$   726\pm   122$&  14.3 \\
12 &$ -72$&$-1$& 0.7456 & 150.556 &$150.584\pm 0.008$&$ 0.452$&  0.046 &$ 0.041\pm0.008$&$   545\pm   112$&  18.2 \\
12 &$ -72$&$ 0$& 0.6672 & 150.737 &$150.764\pm 0.011$&$ 0.468$&  0.060 &$ 0.051\pm0.011$&$   298\pm    79$&  19.8 \\
12 &$ -72$&$ 1$& 0.5911 & 150.880 &$150.951\pm 0.011$&$ 0.485$&  0.074 &$ 0.060\pm0.012$&$   444\pm    99$&  23.6 \\
12 &$ -71$&$-1$& 0.4722 & 151.509 &$151.598\pm 0.010$&$ 0.543$&  0.095 &$ 0.067\pm0.012$&$   650\pm   126$&  27.4 \\
12 &$ -71$&$ 0$& 0.5300 & 151.624 &$151.720\pm 0.009$&$ 0.554$&  0.085 &$ 0.026\pm0.006$&$  1788\pm   329$&  66.9 \\
12 &$ -71$&$ 1$& 0.6122 & 151.772 &$151.861\pm 0.012$&$ 0.567$&  0.070 &$ 0.077\pm0.018$&$   380\pm   113$&  20.2 \\
12 &$ -70$&$ 1$& 0.9339 & 153.268 &$153.289\pm 0.007$&$ 0.697$&  0.012 &$ 0.007\pm0.004$&$   572\pm   100$&  12.3 \\
12 &$ -69$&$ 0$& 0.9717 & 154.828 &$154.849\pm 0.007$&$-0.162$&  0.005 &$ 0.005\pm0.004$&$   587\pm   160$&  11.7 \\
12 &$ -69$&$ 1$& 0.9733 & 155.064 &$155.083\pm 0.007$&$-0.141$&  0.005 &$ 0.006\pm0.004$&$   546\pm    97$&  11.8 \\
\hline
\end{longtable}

\noindent $\zeta$ is derived from the best asymptotic fit; $\nu\ind{as}$ are the asymptotic frequencies, whereas $\nu$ correspond to the observed values; $x=\nu/\Dnu - (\np-\epsp)$ is the reduced frequency; $\Gamma\ind{as}$ are the asymptotic mode widths, whereas $\Gamma$ correspond to the observed values; $H$ are the observed heights, and $R$ is the height-to-background ratio.
}

\begin{table*}
\caption{Oscillation pattern of the RGB star KIC 3955033}\label{table-freq3955033}
\tiny{\begin{tabular}{rrrcccccccc}
  \hline
  $\np$ & $\nm$ & $m$ & $\zeta$&$\nu\ind{as}$ & $\nu$   & $x$ &$\Gamma\ind{as}$& $\Gamma$& $H$  & $R$\\
        &       &     &        &($\mu$Hz)     &($\mu$Hz)&     &  ($\mu$Hz) & ($\mu$Hz)   &(ppm$^2\mu$Hz$^{-1}$)         \\
\hline
 \multicolumn{11}{c}{Radial modes}\\
  \hline
  8 &  &  &  &  84.745 &$ 84.745\pm 0.077$&$ 0.033$&  &$ 0.198\pm0.089$&$  1015\pm   656$&  16.2 \\
 9 &  &  &  &  93.958 &$ 93.762\pm 0.012$&$ 0.010$&  &$ 0.094\pm0.017$&$  3985\pm   872$&  27.3 \\
10 &  &  &  & 102.911 &$103.010\pm 0.013$&$ 0.012$&  &$ 0.138\pm0.027$&$  7641\pm  1745$&  30.2 \\
11 &  &  &  & 112.215 &$112.143\pm 0.009$&$ 0.002$&  &$ 0.067\pm0.012$&$ 18117\pm  3832$& 114.7 \\
12 &  &  &  & 121.869 &$121.474\pm 0.016$&$ 0.013$&  &$ 0.122\pm0.025$&$  2495\pm   637$&  19.6 \\
  \hline
  \multicolumn{11}{c}{Dipole mixed modes}\\
  \hline
 8 &$-144$&$ 1$& 0.8744 &  88.882 &$ 88.894\pm 0.006$&$ 0.483$&  0.019 &$ 0.007\pm0.004$&$  5748\pm  1146$&  15.9 \\
 8 &$-141$&$-1$& 0.8051 &  89.078 &$ 89.075\pm 0.009$&$ 0.503$&  0.029 &$ 0.020\pm0.005$&$  2011\pm   512$&   7.1 \\
 8 &$-142$&$ 0$& 0.7444 &  89.214 &$ 89.205\pm 0.008$&$ 0.517$&  0.038 &$ 0.032\pm0.006$&$  2964\pm   565$&  10.9 \\
 8 &$-143$&$ 1$& 0.6908 &  89.338 &$ 89.338\pm 0.010$&$ 0.531$&  0.046 &$ 0.041\pm0.009$&$  2226\pm   536$&  15.9 \\
 8 &$-140$&$-1$& 0.6640 &  89.507 &$ 89.496\pm 0.007$&$ 0.548$&  0.050 &$ 0.035\pm0.006$&$  6623\pm  1063$&  21.4 \\
 8 &$-141$&$ 0$& 0.6955 &  89.623 &$ 89.602\pm 0.011$&$ 0.560$&  0.046 &$ 0.036\pm0.008$&$  1250\pm   297$&  10.1 \\
 8 &$-142$&$ 1$& 0.7483 &  89.743 &$ 89.725\pm 0.012$&$ 0.573$&  0.038 &$ 0.043\pm0.010$&$  1315\pm   389$&  10.1 \\
 8 &$-139$&$-1$& 0.8432 &  89.969 &$ 89.956\pm 0.009$&$ 0.598$&  0.023 &$ 0.024\pm0.006$&$  2118\pm   541$&   9.8 \\
 8 &$-141$&$ 1$& 0.9111 &  90.251 &$ 90.238\pm 0.009$&$ 0.629$&  0.013 &$ 0.006\pm0.004$&$  2235\pm   601$&   7.4 \\
 9 &$-127$&$-1$& 0.9047 &  97.755 &$ 97.779\pm 0.011$&$ 0.446$&  0.014 &$ 0.023\pm0.007$&$  1041\pm   358$&   9.9 \\
 9 &$-126$&$-1$& 0.6934 &  98.346 &$ 98.392\pm 0.007$&$ 0.512$&  0.046 &$ 0.014\pm0.004$&$ 36038\pm  5534$& 121.7 \\
 9 &$-127$&$ 0$& 0.6800 &  98.375 &$ 98.395\pm 0.007$&$ 0.512$&  0.048 &$ 0.024\pm0.005$&$ 21840\pm  3803$& 121.7 \\
 9 &$-128$&$ 1$& 0.6615 &  98.413 &$ 98.395\pm 0.007$&$ 0.512$&  0.051 &$ 0.027\pm0.005$&$ 19513\pm  3160$& 121.7 \\
 9 &$-125$&$-1$& 0.6969 &  98.825 &$ 98.818\pm 0.007$&$ 0.558$&  0.045 &$ 0.014\pm0.004$&$ 10668\pm  1614$&  47.6 \\
 9 &$-126$&$ 0$& 0.7049 &  98.843 &$ 98.879\pm 0.007$&$ 0.565$&  0.044 &$ 0.028\pm0.006$&$  6104\pm   971$&  27.3 \\
 9 &$-127$&$ 1$& 0.7220 &  98.874 &$ 98.883\pm 0.007$&$ 0.565$&  0.042 &$ 0.031\pm0.006$&$  5427\pm   941$&  27.3 \\
 9 &$-124$&$-1$& 0.9050 &  99.430 &$ 99.435\pm 0.006$&$ 0.625$&  0.014 &$ 0.006\pm0.004$&$  9343\pm  1572$&  29.3 \\
 9 &$-126$&$ 1$& 0.9124 &  99.474 &$ 99.470\pm 0.006$&$ 0.629$&  0.013 &$ 0.012\pm0.004$&$  7355\pm  1118$&  23.3 \\
 9 &$-125$&$ 1$& 0.9647 & 100.170 &$100.166\pm 0.007$&$ 0.704$&  0.005 &$ 0.006\pm0.004$&$  2664\pm   543$&   9.3 \\
10 &$-120$&$ 1$& 0.9883 & 103.940 &$103.940\pm 0.007$&$ 0.113$&  0.002 &$ 0.006\pm0.004$&$  3119\pm   574$&  11.1 \\
10 &$-118$&$-1$& 0.9880 & 104.023 &$104.025\pm 0.010$&$ 0.122$&  0.002 &$ 0.007\pm0.004$&$  2677\pm   836$&  10.2 \\
10 &$-118$&$ 0$& 0.9848 & 104.778 &$104.780\pm 0.011$&$ 0.204$&  0.002 &$ 0.007\pm0.004$&$  2318\pm   827$&   9.3 \\
10 &$-117$&$-1$& 0.9844 & 104.838 &$104.842\pm 0.009$&$ 0.211$&  0.002 &$ 0.006\pm0.004$&$  4417\pm  1446$&  14.7 \\
10 &$-116$&$-1$& 0.9748 & 105.657 &$105.666\pm 0.006$&$ 0.300$&  0.004 &$ 0.005\pm0.004$&$  4718\pm  1166$&  15.8 \\
10 &$-115$&$-1$& 0.9438 & 106.481 &$106.502\pm 0.022$&$ 0.391$&  0.008 &$ 0.019\pm0.011$&$  1349\pm  1022$&  16.4 \\
10 &$-116$&$ 1$& 0.8515 & 107.096 &$107.070\pm 0.008$&$ 0.452$&  0.022 &$ 0.008\pm0.004$&$  1890\pm   368$&  10.1 \\
10 &$-115$&$ 0$& 0.8294 & 107.167 &$107.194\pm 0.011$&$ 0.466$&  0.026 &$ 0.041\pm0.011$&$  1090\pm   354$&   9.5 \\
10 &$-114$&$-1$& 0.8023 & 107.238 &$107.259\pm 0.006$&$ 0.473$&  0.030 &$ 0.016\pm0.004$&$  5430\pm   799$&  21.0 \\
10 &$-115$&$ 1$& 0.5811 & 107.710 &$107.706\pm 0.007$&$ 0.521$&  0.063 &$ 0.019\pm0.005$&$ 22251\pm  3274$&  99.4 \\
10 &$-114$&$ 0$& 0.5753 & 107.757 &$107.776\pm 0.011$&$ 0.529$&  0.064 &$ 0.082\pm0.021$&$  3573\pm  1204$&  70.9 \\
10 &$-113$&$-1$& 0.5782 & 107.819 &$107.818\pm 0.008$&$ 0.533$&  0.063 &$ 0.042\pm0.009$&$  6466\pm  1487$&  71.0 \\
10 &$-114$&$ 1$& 0.7802 & 108.260 &$108.253\pm 0.008$&$ 0.580$&  0.033 &$ 0.018\pm0.005$&$  7228\pm  1753$&  32.4 \\
10 &$-113$&$ 0$& 0.8149 & 108.344 &$108.337\pm 0.009$&$ 0.589$&  0.028 &$ 0.027\pm0.007$&$  3135\pm   809$&  35.8 \\
10 &$-112$&$-1$& 0.8468 & 108.441 &$108.429\pm 0.009$&$ 0.599$&  0.023 &$ 0.029\pm0.007$&$  3843\pm  1006$&  25.7 \\
10 &$-113$&$ 1$& 0.9383 & 109.023 &$109.019\pm 0.006$&$ 0.663$&  0.009 &$ 0.008\pm0.004$&$ 11460\pm  1670$&  41.2 \\
10 &$-112$&$ 1$& 0.9728 & 109.876 &$109.870\pm 0.006$&$ 0.756$&  0.004 &$ 0.006\pm0.004$&$  6439\pm   938$&  23.4 \\
11 &$-104$&$-1$& 0.9239 & 115.807 &$115.818\pm 0.006$&$ 0.400$&  0.011 &$ 0.007\pm0.004$&$  3908\pm   613$&  15.8 \\
11 &$-105$&$ 1$& 0.8432 & 116.274 &$116.279\pm 0.006$&$ 0.450$&  0.023 &$ 0.009\pm0.004$&$  6992\pm  1054$&  26.2 \\
11 &$-104$&$ 0$& 0.7745 & 116.468 &$116.470\pm 0.009$&$ 0.471$&  0.034 &$ 0.033\pm0.008$&$  1247\pm   360$&  18.8 \\
11 &$-103$&$-1$& 0.6842 & 116.648 &$116.670\pm 0.007$&$ 0.492$&  0.047 &$ 0.028\pm0.006$&$  2247\pm   418$&  28.7 \\
11 &$-104$&$ 1$& 0.5371 & 116.961 &$116.976\pm 0.010$&$ 0.525$&  0.069 &$ 0.054\pm0.011$&$  2255\pm   530$&  17.6 \\
11 &$-103$&$ 0$& 0.5477 & 117.092 &$117.077\pm 0.015$&$ 0.536$&  0.068 &$ 0.082\pm0.021$&$  1275\pm   415$&  19.0 \\
11 &$-102$&$-1$& 0.6181 & 117.248 &$117.211\pm 0.010$&$ 0.551$&  0.057 &$ 0.028\pm0.007$&$  1479\pm   387$&  11.9 \\
11 &$-103$&$ 1$& 0.8015 & 117.612 &$117.600\pm 0.007$&$ 0.593$&  0.030 &$ 0.017\pm0.005$&$  2633\pm   464$&  14.5 \\
11 &$-101$&$-1$& 0.9071 & 118.072 &$118.063\pm 0.017$&$ 0.643$&  0.014 &$ 0.031\pm0.013$&$   479\pm   289$&  10.9 \\
\hline
\end{tabular}}

Radial modes and mixed modes identified in KIC 3955033 with a height-to-background ratio $R$ larger than 7. Same caption as Table \ref{table-freq6144777}
\end{table*}

\section{Stars in open clusters}

\begin{figure*}
\includegraphics[width=15cm]{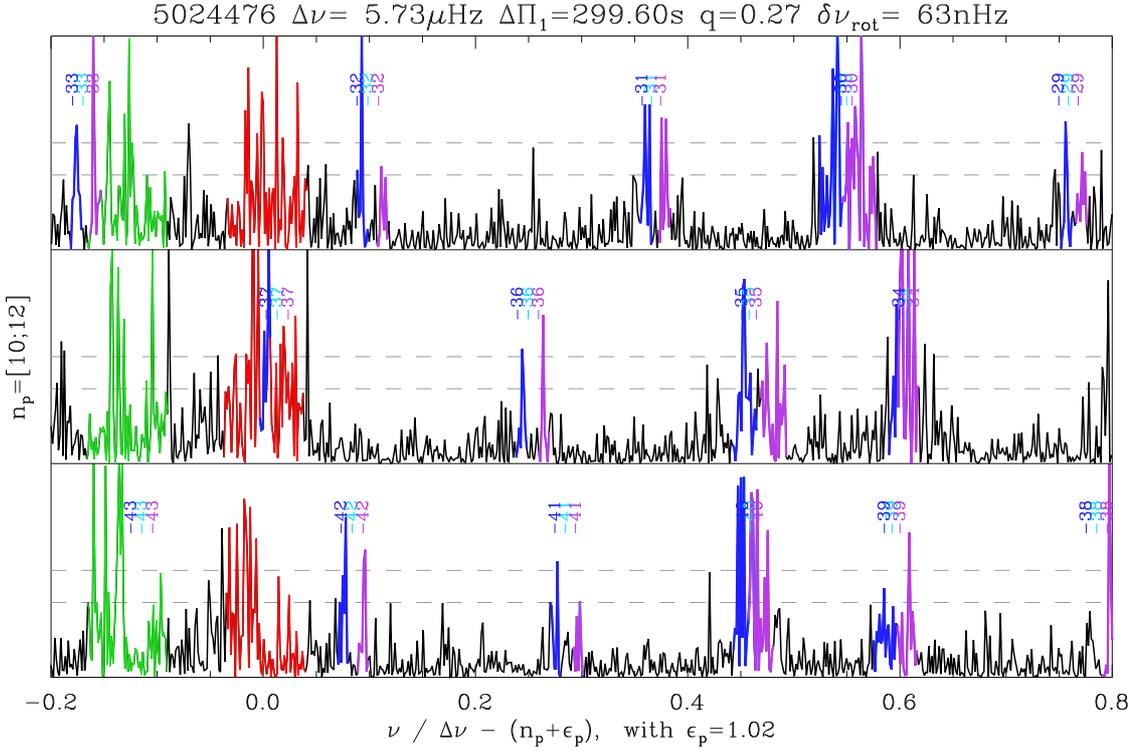}
\caption{Fit of the oscillation pattern of the RGB star KIC 5024476, member of the open cluster NGC 6819. The dim magnitude of the cluster stars explains the low S/R. However, unambiguous doublets are identified all along the mixed-mode spectrum;  $m=0$ modes are mostly absent and $|m|=1$ modes dominate the mixed-mode spectrum, so that a nearly pole-on inclination is not possible. Same style as Fig.~\ref{fig-fit-6144777}.}\label{fig-fit-cluster1}
\end{figure*}

\begin{figure*}
\includegraphics[width=15cm]{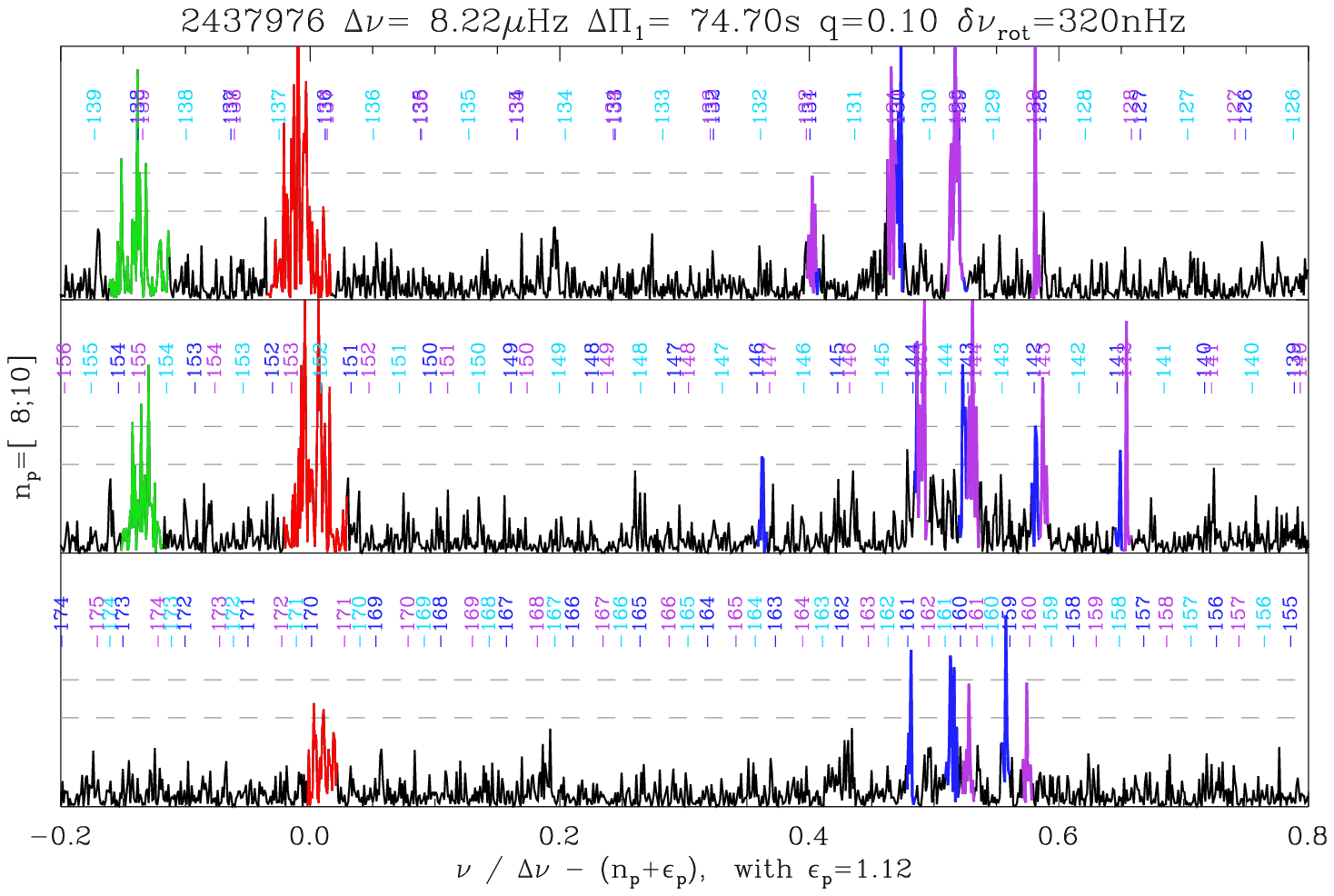}
\caption{Fit of the oscillation pattern of the RGB star KIC 2437976, member of the open cluster NGC 6791. The dim magnitude of the cluster stars explains the low S/R. The identification at radial order 9, supported by the radial orders 8 and 10, is unambiguously conclusive: $m=0$ modes are mostly absent and $|m|=1$ modes dominate the mixed-mode spectrum, so that a pole-on inclination is not possible. Same style as Fig.~\ref{fig-fit-6144777}.}\label{fig-fit-cluster2}
\end{figure*}

All stars studied by \cite{2017NatAs...1E..64C} were investigated. The fitting process is challenging, due to the dim magnitudes of such dim stars in open clusters. However, the combination of all pressure radial orders near $\numax$ provides in most cases an unambiguous fit, and at least a few mixed-mode radial orders provide clear splittings.\\
- Figure~\ref{fig-fit-cluster1} provides the asymptotic fit of KIC 5024476, member of the open cluster NGC 6819 observed by \Kepler. We note that $m=\pm 1$ modes are clearly identified and derive a stellar inclination $i = 79 \pm 11^\circ$ for this star. This result is in disagreement with \cite{2017NatAs...1E..64C} who found an inclination  $i = 20 \pm 7^\circ$.\\
- Similar conclusions are reached for KIC 2437976 (Fig.~\ref{fig-fit-cluster2}), member of the open cluster NGC 6791. \cite{2017NatAs...1E..64C} found an inclination  $i = 0 \pm 10^\circ$, despite the fact $|m|=1$ modes are clearly identified and indicate $i = 76 \pm 14^\circ$. \\

These stars are representative of the whole data set treated by \cite{2017NatAs...1E..64C}: the inability of the fitting process to identify thin short-lived mixed modes translates into the identification of a single broad $m=0$ peak. In such cases, stellar inclinations derived from the Bayesian fits are necessarily underestimated and biased toward low values.

\begin{table*}
\caption{Oscillation pattern of the red-clump star KIC 5024476 in NGC 6819\label{table-freq5024476}}
\begin{tabular}{rrrcccccccc}
\hline
  $\np$ & $\nm$ & $m$ & $\zeta$&$\nu\ind{as}$ & $\nu$   & $x$ &$\Gamma\ind{as}$& $\Gamma$& $H$  & $R$\\
        &       &     &        &($\mu$Hz)     &($\mu$Hz)&     &  ($\mu$Hz) & ($\mu$Hz)   &(ppm$^2\mu$Hz$^{-1}$)         \\
\hline
 \multicolumn{11}{c}{Radial modes}\\
  \hline
10 &  &  &  &  63.335 &$ 63.096\pm 0.174$&$-0.058$&  &$ 0.501\pm0.216$&$   698\pm   429$&   9.9 \\
11 &  &  &  &  69.039 &$ 69.039\pm 0.055$&$-0.021$&  &$ 0.350\pm0.102$&$  1365\pm   536$&   9.7 \\
12 &  &  &  &  74.941 &$ 74.941\pm 0.039$&$ 0.009$&  &$ 0.231\pm0.067$&$  1158\pm   452$&  35.6 \\
  \hline
  \multicolumn{11}{c}{Dipole mixed modes}\\
  \hline
10 &$ -43$&$-1$& 0.9495 &  63.873 &$ 63.870\pm 0.009$&$ 0.077$&  0.013 &$ 0.019\pm0.006$&$  1887\pm   532$&  16.0 \\
10 &$ -43$&$ 1$& 0.9481 &  63.979 &$ 63.976\pm 0.007$&$ 0.095$&  0.013 &$ 0.009\pm0.004$&$  3002\pm   632$&  12.7 \\
10 &$ -42$&$-1$& 0.9006 &  65.035 &$ 65.005\pm 0.015$&$ 0.275$&  0.025 &$ 0.045\pm0.012$&$   497\pm   175$&  11.6 \\
10 &$ -42$&$ 0$& 0.8955 &  65.086 &$ 65.131\pm 0.012$&$ 0.297$&  0.026 &$ 0.028\pm0.008$&$   576\pm   191$&   7.6 \\
10 &$ -42$&$ 1$& 0.8903 &  65.136 &$ 65.131\pm 0.011$&$ 0.297$&  0.027 &$ 0.027\pm0.008$&$   589\pm   186$&   7.6 \\
10 &$ -41$&$-1$& 0.6178 &  66.050 &$ 66.050\pm 0.013$&$ 0.457$&  0.096 &$ 0.095\pm0.020$&$  1859\pm   503$&  20.0 \\
10 &$ -41$&$ 0$& 0.6011 &  66.085 &$ 66.061\pm 0.013$&$ 0.459$&  0.100 &$ 0.099\pm0.022$&$  1959\pm   563$&  20.0 \\
10 &$ -41$&$ 1$& 0.5866 &  66.117 &$ 66.083\pm 0.014$&$ 0.463$&  0.103 &$ 0.106\pm0.025$&$  1783\pm   534$&  18.7 \\
10 &$ -40$&$-1$& 0.6453 &  66.784 &$ 66.788\pm 0.017$&$ 0.586$&  0.089 &$ 0.085\pm0.020$&$   800\pm   240$&   8.9 \\
10 &$ -40$&$ 1$& 0.6794 &  66.858 &$ 66.897\pm 0.020$&$ 0.605$&  0.080 &$ 0.097\pm0.023$&$   632\pm   197$&  14.4 \\
10 &$ -39$&$ 1$& 0.9132 &  67.985 &$ 68.003\pm 0.007$&$ 0.798$&  0.022 &$ 0.009\pm0.004$&$  4805\pm   844$&  23.5 \\
11 &$ -37$&$-1$& 0.9084 &  70.563 &$ 70.562\pm 0.007$&$ 0.244$&  0.023 &$ 0.009\pm0.004$&$  2295\pm   521$&  11.4 \\
11 &$ -37$&$ 1$& 0.9003 &  70.665 &$ 70.672\pm 0.006$&$ 0.264$&  0.025 &$ 0.008\pm0.004$&$  2474\pm   398$&  14.8 \\
11 &$ -36$&$-1$& 0.6140 &  71.776 &$ 71.786\pm 0.015$&$ 0.458$&  0.097 &$ 0.103\pm0.022$&$  1228\pm   333$&  18.4 \\
11 &$ -36$&$ 1$& 0.5811 &  71.843 &$ 71.884\pm 0.019$&$ 0.475$&  0.105 &$ 0.127\pm0.030$&$   854\pm   264$&  12.1 \\
11 &$ -35$&$-1$& 0.6100 &  72.588 &$ 72.615\pm 0.010$&$ 0.603$&  0.097 &$ 0.072\pm0.013$&$  2715\pm   560$&  29.9 \\
11 &$ -35$&$ 1$& 0.6455 &  72.658 &$ 72.625\pm 0.011$&$ 0.605$&  0.089 &$ 0.078\pm0.015$&$  2287\pm   527$&  29.9 \\
11 &$ -34$&$-1$& 0.9051 &  73.870 &$ 73.881\pm 0.007$&$-0.176$&  0.024 &$ 0.016\pm0.005$&$  2320\pm   417$&  12.5 \\
11 &$ -34$&$ 1$& 0.9111 &  73.972 &$ 73.977\pm 0.007$&$-0.160$&  0.022 &$ 0.008\pm0.004$&$  4124\pm   750$&  22.2 \\
12 &$ -33$&$-1$& 0.9314 &  75.420 &$ 75.422\pm 0.006$&$ 0.093$&  0.017 &$ 0.007\pm0.004$&$  5610\pm   987$&  31.2 \\
12 &$ -33$&$ 1$& 0.9296 &  75.524 &$ 75.530\pm 0.008$&$ 0.112$&  0.018 &$ 0.008\pm0.004$&$  1498\pm   379$&   8.3 \\
12 &$ -32$&$-1$& 0.8053 &  76.972 &$ 76.954\pm 0.008$&$ 0.360$&  0.049 &$ 0.024\pm0.006$&$  1836\pm   391$&  14.5 \\
12 &$ -32$&$ 1$& 0.7828 &  77.061 &$ 77.060\pm 0.009$&$ 0.379$&  0.054 &$ 0.034\pm0.007$&$  1103\pm   222$&  13.2 \\
12 &$ -31$&$-1$& 0.4338 &  78.016 &$ 78.029\pm 0.017$&$ 0.548$&  0.142 &$ 0.156\pm0.035$&$  1362\pm   387$&  32.1 \\
12 &$ -31$&$ 0$& 0.4373 &  78.041 &$ 78.045\pm 0.018$&$ 0.550$&  0.141 &$ 0.160\pm0.036$&$  1326\pm   379$&  16.4 \\
12 &$ -31$&$ 1$& 0.4416 &  78.066 &$ 78.062\pm 0.017$&$ 0.553$&  0.140 &$ 0.154\pm0.034$&$  1354\pm   382$&  14.3 \\
12 &$ -30$&$-1$& 0.8397 &  79.185 &$ 79.226\pm 0.008$&$ 0.757$&  0.040 &$ 0.013\pm0.004$&$  2127\pm   377$&  12.8 \\
12 &$ -30$&$ 1$& 0.8538 &  79.280 &$ 79.310\pm 0.009$&$ 0.771$&  0.037 &$ 0.027\pm0.006$&$   977\pm   236$&   9.7 \\
\hline
\end{tabular}

\noindent Same caption as Table \ref{table-freq6144777}
\end{table*}

\begin{table*}
\caption{Oscillation pattern of the RGB star KIC 2437976 in NGC 6791 \label{table-freq2437976}}
\begin{tabular}{rrrcccccccc}
\hline
  $\np$ & $\nm$ & $m$ & $\zeta$&$\nu\ind{as}$ & $\nu$   & $x$ &$\Gamma\ind{as}$& $\Gamma$& $H$  & $R$\\
        &       &     &        &($\mu$Hz)     &($\mu$Hz)&     &  ($\mu$Hz) & ($\mu$Hz)   &(ppm$^2\mu$Hz$^{-1}$)         \\
\hline
 \multicolumn{11}{c}{Radial modes}\\
  \hline
 8 &  &  &  &  75.045 &$ 75.045\pm 0.041$&$ 0.013$&  &$ 0.106\pm0.042$&$  6520\pm  3676$&   7.8 \\
 9 &  &  &  &  83.284 &$ 83.172\pm 0.024$&$ 0.001$&  &$ 0.153\pm0.036$&$ 15645\pm  4765$&  31.3 \\
10 &  &  &  &  91.382 &$ 91.331\pm 0.013$&$-0.006$&  &$ 0.100\pm0.017$&$ 22777\pm  4443$&  31.1 \\
  \hline
  \multicolumn{11}{c}{Dipole mixed modes}\\
  \hline
 8 &$-161$&$-1$& 0.8533 &  78.887 &$ 78.893\pm 0.008$&$ 0.481$&  0.026 &$ 0.022\pm0.006$&$ 10581\pm  2376$&  12.3 \\
 8 &$-160$&$-1$& 0.6438 &  79.234 &$ 79.170\pm 0.010$&$ 0.514$&  0.064 &$ 0.038\pm0.007$&$ 14647\pm  2841$&  11.9 \\
 8 &$-161$&$ 1$& 0.6511 &  79.343 &$ 79.276\pm 0.012$&$ 0.527$&  0.063 &$ 0.031\pm0.007$&$ 10930\pm  2616$&   9.6 \\
 8 &$-159$&$-1$& 0.7884 &  79.561 &$ 79.519\pm 0.008$&$ 0.557$&  0.038 &$ 0.026\pm0.006$&$ 14986\pm  2978$&  15.1 \\
 8 &$-160$&$ 1$& 0.8566 &  79.691 &$ 79.671\pm 0.009$&$ 0.575$&  0.026 &$ 0.023\pm0.006$&$  6508\pm  1802$&   9.8 \\
 9 &$-146$&$-1$& 0.9726 &  86.115 &$ 86.141\pm 0.010$&$ 0.362$&  0.005 &$ 0.009\pm0.004$&$ 16564\pm  5717$&   7.6 \\
 9 &$-144$&$-1$& 0.7541 &  87.141 &$ 87.185\pm 0.010$&$ 0.489$&  0.044 &$ 0.054\pm0.013$&$ 20608\pm  6137$&  33.9 \\
 9 &$-145$&$ 1$& 0.7180 &  87.193 &$ 87.191\pm 0.009$&$ 0.490$&  0.051 &$ 0.049\pm0.011$&$ 23464\pm  6034$&  33.9 \\
 9 &$-143$&$-1$& 0.6110 &  87.503 &$ 87.500\pm 0.011$&$ 0.528$&  0.070 &$ 0.067\pm0.014$&$ 19703\pm  4780$&  29.1 \\
 9 &$-144$&$ 1$& 0.6404 &  87.550 &$ 87.517\pm 0.009$&$ 0.530$&  0.065 &$ 0.050\pm0.009$&$ 26627\pm  5525$&  29.1 \\
 9 &$-142$&$-1$& 0.8736 &  87.940 &$ 87.943\pm 0.007$&$ 0.582$&  0.023 &$ 0.009\pm0.004$&$ 21699\pm  5272$&  10.0 \\
 9 &$-143$&$ 1$& 0.8929 &  88.001 &$ 87.989\pm 0.006$&$ 0.587$&  0.019 &$ 0.009\pm0.004$&$ 29926\pm  5102$&  13.8 \\
 9 &$-141$&$-1$& 0.9588 &  88.486 &$ 88.498\pm 0.007$&$ 0.649$&  0.007 &$ 0.007\pm0.004$&$ 17180\pm  3838$&   8.1 \\
 9 &$-142$&$ 1$& 0.9622 &  88.537 &$ 88.540\pm 0.006$&$ 0.654$&  0.007 &$ 0.006\pm0.004$&$ 38360\pm  5733$&  18.3 \\
10 &$-132$&$ 1$& 0.9463 &  94.663 &$ 94.689\pm 0.007$&$ 0.402$&  0.010 &$ 0.006\pm0.004$&$ 19058\pm  3790$&   9.7 \\
10 &$-131$&$-1$& 0.9440 &  94.692 &$ 94.708\pm 0.008$&$ 0.405$&  0.010 &$ 0.005\pm0.004$&$ 15841\pm  4952$&   7.6 \\
10 &$-131$&$ 1$& 0.7969 &  95.259 &$ 95.216\pm 0.009$&$ 0.466$&  0.037 &$ 0.034\pm0.008$&$ 21679\pm  5266$&  18.4 \\
10 &$-130$&$-1$& 0.7827 &  95.284 &$ 95.257\pm 0.009$&$ 0.471$&  0.039 &$ 0.033\pm0.008$&$ 19865\pm  4877$&  20.4 \\
10 &$-130$&$ 1$& 0.5518 &  95.692 &$ 95.630\pm 0.008$&$ 0.517$&  0.081 &$ 0.025\pm0.005$&$ 42908\pm  7274$&  26.4 \\
10 &$-129$&$-1$& 0.5572 &  95.709 &$ 95.633\pm 0.008$&$ 0.517$&  0.080 &$ 0.027\pm0.006$&$ 39045\pm  6386$&  26.4 \\
10 &$-129$&$ 1$& 0.8453 &  96.161 &$ 96.162\pm 0.007$&$ 0.582$&  0.028 &$ 0.009\pm0.004$&$ 29045\pm  5946$&  21.6 \\
10 &$-125$&$ 0$& 0.9891 &  98.575 &$ 98.551\pm 0.008$&$-0.128$&  0.002 &$ 0.004\pm0.004$&$ 15436\pm 11044$&   7.5 \\
11 &$-118$&$-1$& 0.8082 & 103.469 &$103.439\pm 0.010$&$ 0.467$&  0.034 &$ 0.023\pm0.006$&$  6844\pm  1785$&   9.8 \\
12 &$-108$&$ 1$& 0.7958 & 111.880 &$111.821\pm 0.012$&$ 0.487$&  0.037 &$ 0.022\pm0.006$&$  5310\pm  1445$&   7.9 \\
12 &$-105$&$-1$& 0.9174 & 113.342 &$113.329\pm 0.007$&$ 0.670$&  0.015 &$ 0.006\pm0.004$&$ 16746\pm  4518$&   8.6 \\
\hline
\end{tabular}

\noindent Same caption as Table \ref{table-freq6144777}
\end{table*}

\end{appendix}

\end{document}